# Targets of drugs are generally, and targets of drugs having side effects are specifically good spreaders of human interactome perturbations

**Áron R. Perez-Lopez[1,*], Kristóf Z. Szalay[1], Dénes Türei[2,+], Dezső Módos[2,3], Katalin Lenti[3], Tamás Korcsmáros[2,4,5] and Peter Csermely[1,#]**

**1** Department of Medical Chemistry, Semmelweis University, P.O. Box 260, H-1444 Budapest 8, Hungary; **2** Department of Genetics, Eötvös Loránd University, Pázmány P. s. 1C, H-1117 Budapest, Hungary; **3** Department of Morphology and Physiology, Faculty of Health Sciences, Semmelweis University, Vas u. 17, H-1088 Budapest, Hungary; **4** TGAC, The Genome Analysis Centre, Norwich, UK; **5** Gut Health and Food Safety Programme, Institute of Food Research, Norwich, UK

*Áron R. Perez Lopez is a high school student of the Apáczai Csere János High School of the Eötvös Loránd University, Papnövelde u. 4-6, H-1053 Budapest, Hungary; +current address: European Bioinformatics Institute (EMBL-EBI), Wellcome Trust Genome Campus, Cambridge CB10 1SD, UK
#Corresponding author, E-mail: csermely.peter@med.semmelweis-univ.hu

## Abstract

Network-based methods are playing an increasingly important role in drug design. Our main question in this paper was whether the efficiency of drug target proteins to spread perturbations in the human interactome is larger if the binding drugs have side effects, as compared to those which have no reported side effects. Our results showed that in general, drug targets were better spreaders of perturbations than non-target proteins, and in particular, targets of drugs with side effects were also better spreaders of perturbations than targets of drugs having no reported side effects in human protein-protein interaction networks. Colorectal cancer-related proteins were good spreaders and had a high centrality, while type 2 diabetes-related proteins showed an average spreading efficiency and had an average centrality in the human interactome. Moreover, the interactome-distance between drug targets and disease-related proteins was higher in diabetes than in colorectal cancer. Our results may help a better understanding of the network position and dynamics of drug targets and disease-related proteins, and may contribute to develop additional, network-based tests to increase the potential safety of drug candidates.

## Introduction

Due to the "curse of attrition" drug side effects are subjects of increasing concerns[1-4]. In recent years a growing number of side effect databases helped pharmacovigilance efforts[2,5-10]. In addition, the prediction of drug side effects was a subject of several excellent network studies. These contributions constructed and analyzed drug—side effect networks[1,8,11], side effect similarity-based drug—drug networks[12-14], drug target—side effect networks (including correlated drug binding profiles and side effect profiles and protein domain networks)[3,5,7,15,16], as well as drug—side effect—biological pathway multi-layer networks[9,10,17,18].

Parallel with the sequencing of the human genome, the pharmaceutical industry increasingly turned towards rational drug design, where drug target candidates are selected on the basis of known disease-related genes. In recent years, however, it became apparent that drug action often extends beyond its primary target, and also affects the neighbourhood of the primary target in molecular networks[4,19-23]. The influence on network neighbourhood can be efficiently modelled as a spreading process. Indeed, network spreading efficiency became increasingly used to characterize the dynamics of a wide variety of networks, such as the propagation of infections and computer viruses[24-26], as well as the spread of information, innovations and social influence[27-30]. Long-range spread of conformational changes *via* protein-protein interaction networks is supported by several pieces of experimental evidence[31,32]. Moreover, recent studies extended the use of information-spread to molecular networks highlighting the usefulness of this approach in finding key amino acids of protein structure networks, biologically relevant changes of cellular functions upon stress, reprogramming biological networks, and uncovering the attractor changes in malignant transformation[33-36]. However, network spreading efficiency has been used to characterize drug targets neither in general, nor restricted to targets of drugs having side effects.

In this study we investigated, whether the efficiency of drug target proteins to spread perturbations in the human interactome is larger, if drugs targeting them have side effects, as compared to the spreading efficiency of targets of those drugs, which have no reported side effects. Encouraged by our findings that drug targets in general, and targets of drugs having side effects in particular, spread perturbation better in the human interactome than other proteins, we specifically examined two diseases, colorectal cancer and diabetes. These two, wide-spread diseases were selected, since they represent target groups of different drug design strategies[4], and they had been the subjects of several former network-related studies[37-45]. We found that colorectal cancer-related proteins were good spreaders and had a high centrality in the human protein-protein interaction network. On the contrary, type 2 diabetes-related proteins showed an average spreading efficiency, and had an average centrality. Additionally, network shortest path (geodesic distance) between drug targets and disease-related proteins was higher in diabetes than in colorectal cancer. Our results give novel details on the network topology and dynamics of disease-related and drug target proteins, and may initiate the development of novel, network-based pharmacovigilance methods increasing the potential safety of drug candidates.

# Results

## Targets of drugs with side effects spread perturbations better in the human interactome than targets of drugs without side effects

The initial working hypothesis of our research was that drugs having protein targets that better propagate changes in the human interactome may have a higher probability of causing side effects. This hypothesis is in agreement with earlier findings showing that the interactome neighbourhood contributed to drug side-effect similarity[20]. In order to test our hypothesis, we compared the propagation of perturbations started from drug targets with and without known side effect, as well as that of non-target proteins in the human protein-protein interaction network using the Turbine network dynamics software package developed earlier in our group[35].

To compare the spreading efficiency of drug target proteins with and without side effects we ran a series of perturbation simulations on the human interactome using the Turbine programme[35]. We assembled a human interactome containing 12,439 proteins and 174,666 edges using the STRING database[46], out of which 1,726 were target proteins of 3,626 human drugs obtained from the DrugBank database[47] and a total of 99,423 drug-side effect pairs from the SIDER database[2] were analysed as described in Methods in detail. Simulations were based on the communicating vessels network dynamics model tested earlier[35], where changes from one protein to its neighbours 'flow' in proportion with the energy differences between the 'source' and the 'target' proteins. We examined a total of 495 target proteins of 597 drugs (Suppl. Table 1), which were reported to have side effects according to the SIDER database[2]. As control groups, we have also examined the 1,231 target proteins of the remaining 3,029 drugs having no reported side effects in the SIDER database[2], as well as the remaining 10,713 proteins in our human interactome, which were not listed as drug targets in DrugBank[47]. For each selected protein target we calculated the silencing time, which is the number of time steps in the simulation needed for the initial perturbation to disappear completely due to dissipation. Small silencing time values were shown to be an efficient measure of large spreading efficiency of network nodes earlier[35], since in this case the initial perturbation efficiently spreads in the network and it becomes dissipated fast.

Fig. 1 shows the cumulative distribution of the normalized number of proteins having an increasing silencing time (thus decreasing perturbation efficiency). Targets of drugs with side effects had a significantly larger proportion of small silencing times (i.e. large spreading efficiency) than targets of drugs having no side effects (Mann-Whitney-Wilcoxon test, $p=1.677e-5$). Similarly, the proportion of targets of drugs without side effects having a small silencing time (i.e. large spreading efficiency) was significantly larger than that of human interactome proteins, which have not been reported as drug targets in DrugBank[47] (Mann-Whitney-Wilcoxon test, $p=2.2e-16$). Thus targets of drugs with side effects were found to be better spreaders of perturbations than targets of drugs having no reported side effects. Importantly, drug targets were also better spreaders of perturbations than non-target proteins.

Simulations shown on Fig. 1 were run with a starting energy of 1,000 units and a dissipation value of 5 units. Being curious whether our result is robust for the variations of simulation parameters, we repeated these simulations using a starting energy of 10,000 and a dissipation of 1 or 5 units. Under these conditions we obtained very similar results (Suppl. Figs. 1 and 2) to those shown on Fig. 1. When we split the starting energy of 1,000 units equally among targets of multi-target drugs instead of examining each target protein alone as the source of perturbations, we were able to reproduce the same pattern (Suppl. Fig. 3) as that of Fig. 1. Furthermore, to test the robustness of the results against the choice of protein-protein interaction network, we randomly deleted 50% of the 12,439 proteins in our human interactome. Examining the spreading efficiency in the giant component of this truncated interactome we obtained very similar results (Suppl. Fig. 4) to those shown in Fig. 1.

Next we were curious whether the larger spreading efficiency of drug targets with side effects, as compared to drug targets without side effects or proteins having no reported drugs bound to them, is also shown by examining perturbation reach values. Perturbation reach values show the number of proteins, which received the perturbation from the initial perturbation source protein until the perturbation was dissipated from the system. Small perturbation reach values were shown to characterize small spreading efficiency in earlier studies[35], since in this case the original perturbation reached only a small number of proteins before it became dissipated. Targets of drugs with side effects had a significantly smaller proportion of small perturbation reach values (i.e. small spreading efficiency) than that of targets of drugs having no side effects (Mann-Whitney-Wilcoxon test, p=1.663e-5; Suppl. Fig. 5). Similarly, the proportion of targets of drugs without side effects having a small perturbation reach value (i.e. small spreading efficiency) was significantly smaller than that of human interactome proteins, which have not been reported as drug targets in DrugBank[47] (Mann-Whitney-Wilcoxon test, p=2.2e-16; Suppl. Fig. 5). Using a starting energy of 10,000 but a dissipation of 1 instead of 5 units, or splitting this starting energy equally among targets of multi-target drugs, we obtained very similar results (Suppl. Figs. 6 and 7). These studies confirmed that drug targets are better spreaders of perturbations than non-target proteins, and also that targets of drugs with side effects are better spreaders of perturbations than targets of drugs having no reported side effects.

A qualitatively similar picture emerged, when we examined the spreading efficiency of target proteins of drugs against two diseases, colorectal cancer and type 2 diabetes (Suppl. Tables 2-6). We chose these two diseases, because they represent very well the target groups of different drug design strategies[4], and they had been the subjects of several former network-related studies[37-45]. Drug targets of both diseases were found to be better spreaders of perturbations than non-target proteins (Suppl. Fig. 8; p=3.367e-5 and p=5.88e-5 for colorectal cancer and diabetes, respectively). There was a tendency showing that targets of drugs with side effects were better spreaders of perturbations than targets of drugs having no reported side effects both in colorectal cancer and in diabetes. However, due to the low number of identified drug targets having side effects (3 and 25, respectively), these latter differences were not statistically significant (p=1 and p=0.2593, respectively).

**Colorectal cancer-related proteins are good spreaders of perturbations and have a high centrality, while type-2 diabetes-related proteins show an average spreading efficiency and average centrality**

Very importantly, a rather interesting difference emerged, when we examined the spreading efficiency of proteins related to colorectal cancer and diabetes. Mutated genes and their corresponding proteins in colorectal cancer and in type-2 diabetes were obtained from the Cancer Gene Census database[48] (Suppl. Table 7) and from the article of Parchwani et al.[49] (Suppl. Table 8), respectively. In case of colorectal cancer, disease-associated proteins were found to be significantly better spreaders than the residual proteins of the human interactome. On the contrary, diabetes-related proteins showed indistinguishable spreading properties to the rest of human proteins, which were not associated with the onset of diabetes (Fig. 2). To test the robustness of the results against the choice of protein-protein interaction network, we randomly deleted 50% of the 12,439 proteins in our human interactome. Here again, colorectal cancer-associated proteins were found to be significantly better spreaders than the residual proteins of the human interactome (data not shown; p=0.00021 in Mann-Whitney test) and spreading efficiency of diabetes-related proteins showed no significant difference as compared to the rest of human proteins (data not shown; p=0.095 in Mann-Whitney test).

These findings are in agreement with earlier results showing that cancer-associated proteins are enriched in proteins having a high centrality in the human interactome[37,38,40,42-45]. Indeed, in our human interactome, cancer-related proteins had a significantly higher degree, closeness and betweenness centralities than diabetes-related proteins, having a 9.6-, 1.2- and 54-fold increase, respectively (Table 1). In agreement with their similar silencing time values (Suppl. Fig. 8), drug targets without or with side effects showed no significant centrality differences in the human interactome (Suppl. Table 9).

**The interactome distance between drug targets and disease-related proteins is higher in diabetes than in colorectal cancer**

Encouraged by the results showing an increased centrality of cancer-related, but not of diabetes-related proteins in the human interactome, we examined the interactome geodesic distance (i.e. shortest path) between drug targets and disease related proteins in both diseases using the neighbourhood matrices of related proteins. Our data show that the geodesic distance in the human interactome between drug targets and disease-related proteins is significantly larger in case of type-2 diabetes than in colorectal cancer (targets without side effects: p=1.062e-5; targets with side effects: p=5.441e-3). (Table 2; Suppl. Tables 10-13 and Suppl. Fig. 9) This finding is supported by the visual representation of the human sub-interactome of drug target and disease-related proteins of these two diseases (Suppl. Fig. 10), where drug targets and disease-related proteins of colorectal cancer are intertwined, while these two groups of proteins remain rather separated in type-2 diabetes. This observation is further substantiated by the fact, that only 1 of the 18 colorectal cancer-related proteins (6%) is not connected to the giant component of the sub-interactome, while 10 of the 14 diabetes-related proteins (71%) are missing from the same giant component (Suppl. Fig. 10).

## Discussion

The most important finding of our study is that 1.) drug targets are better spreaders of perturbations in the human interactome than non-target proteins in general; and in particular, 2.) targets of drugs with side effects are also better spreaders of perturbations than targets of drugs having no reported side effects (Fig. 1). These findings were robust, since they could be reproduced when we used different perturbation parameters (Suppl. Figs. 1, 2 and 3), different measures of perturbation spread (Suppl. Figs. 5, 6 and 7), and reduced the size (coverage) of the human interactome to half of the original (Suppl. Fig. 4). These results are in agreement with those of a previous study showing that the interactome neighbourhood contributed to side-effect similarity[20].

Importantly, colorectal cancer-related proteins are good spreaders of perturbations and had a high centrality, while type-2 diabetes-related proteins showed an average spreading efficiency and had an average centrality in the human interactome (Fig. 2 and Table 1). These findings are in agreement with earlier results showing that cancer-associated proteins are enriched in hubs, bottlenecks and bridges all having a high centrality in the human interactome[37,38,40,42-45].

Furthermore, the interactome-distance between drug targets and disease-related proteins was higher in diabetes than in colorectal cancer (Table 2; Suppl. Tables 10-13 and Suppl. Fig. 9). This finding is in agreement with both the results of previous studies and intuitive insights on the classification of drug target strategies[4]. Most drug targets are 3 or 4 steps away in the human interactome from proteins involved in the same disease[50]. Moreover, cancer-related and metabolic disease-related proteins were shown to have an average network distance to the related drug targets of 2.3 and ~5 network edges, which are smaller and higher than the most abundant distance values, respectively, forming the two extremes of the distance-spectrum[50]. The former value is in the range we found in our study (Table 2). The latter value of a disease group containing diabetes is much larger than that related to cancer, which is again in agreement with our findings. As a general trend, rapidly proliferating cells, like those in cancer, are attacked at their central proteins, while differentiated cells, such as those involved in type-2 diabetes, are attacked at the neighbours of central proteins[4]. These assumptions are also in agreement with a smaller network distance of centrally positioned cancer-related proteins from centrally positioned cancer drug targets than the distance between the more peripheral diabetes-related proteins and drug targets.

Analysis of perturbation spread in molecular networks may be used to develop additional, network-based tests to increase the potential safety of drug candidates. Assessment of perturbation spread in weighted networks (where the edges are weighted according to the abundance of their end-node proteins of relevant tissues, e.g. the endothelial cell in colorectal cancer, as well as hepatocyte and myocyte in diabetes, as described in our earlier study for the yeast interactome[51]), directed networks (such as signalling networks[4,52]), or networks considering the subcellular localization of participating proteins[53], as well as using quantitative measures of side-effect severity and abundance may provide additional information and will be subjects of later studies.

In summary, our results contributed to a better understanding of the network position and dynamics of disease-related and drug target proteins. The findings may help the future development of novel, network-based pharmacovigilance methods increasing the potential safety of drug candidates.

## Methods

**Construction of the human protein-protein interaction network**

In this paper, we examined the propagation of perturbations in the human protein-protein interaction network (interactome). The choice of this type of network was driven by the fact that it contains the most proteins and the greatest number of connections (as opposed to signalling networks or regulatory networks). Human interactome data were downloaded from the STRING database[46] on 8 February, 2013. STRING contains interaction data based on a vast number of data collection principles. We have only used manually collected ('database' column) or experimental ('experiments' column) data having higher reliability than e.g. predicted data. Only human protein-protein interactions were included in the interactome. In order to facilitate the comparison with drug targets, the STRING Ensemble Protein ID (ENSP) protein codes were translated to UniProt ID[54] using the UniProt translator. From the original 13,484 ENSP IDs we managed to translate 12,493 to UniProt IDs, but only 12,439 proteins were connected to other proteins. The database contained a total of 377,920 human protein-protein interactions, out of which 350,528 remained after translating the protein IDs to UniProt IDs using the UniProt translator, which were further reduced to 174,666 after eliminating multiple links and loops (self-links). The original STRING database also contained edge weights indicating the reliability of data. Since we only worked with manually collected and experimental data, our interactome contained no edge weights.

**Measurement of the propagation of perturbations in the human interactome**

The propagation of perturbations in the human interactome was measured with the network perturbation analysis software for simulating network dynamics called Turbine[35]. For the simulation experiments we chose the software's communicating vessels model[35], where changes from one protein to its neighbours 'flow' in proportion with the energy differences between the 'source' and the 'target' proteins. The communicating vessels model[35] contains a starting energy ($E$) and a dissipation parameter ($D$), where the starting energy is distributed equally among the proteins of the human interactome specified at the individual simulations, while in each step of the simulation the program subtracts $D$ units of energy from each protein of the interactome. In most simulations $E$ and $D$ were set to 1000 and 5 units, respectively. Having these starting energy and dissipation parameters it was possible to trace the propagation of perturbations in the network rather easily. However, all the key simulations were also examined using different $E$ and $D$ values to examine the robustness of the results. To characterise the propagation efficiency of the starting node(s), the measure of silencing time[35] was used, which is the time elapsed from the start of the simulation until the energy of all nodes reaches the minimum threshold of less than 1 unit. We also calculated perturbation reach values[35], which show the number of proteins receiving the perturbation from the initial perturbation source protein until the perturbation was dissipated from the system.

**Characterisation of drug side effects**

Drug side effects were collected from the SIDER database[2]. This database contains information about drug side effects and their frequencies from public documentation and package inserts, with the help of drug labels and terms from MedDRA (Medical Dictionary for Regulatory Activities). SIDER data were downloaded from the version of 17 October, 2012. This version of the SIDER database[2] contained 996 drugs, 4,192 unique side effects and 215,850 drug-side effect pairs. After eliminating the duplicates, 99,423 drug-side effect pairs remained. In order to be able to compare data, we converted drug IDs in the SIDER database[2] into IDs of the DrugBank database[47] by matching the drug names.

**Characterisation of drug targets**

We collected drug targets from the DrugBank database[47] version last updated on 10 February, 2013. The XML version of the database was used, including the drug names, indications and target list. The proteins in the target list were identified by their UniProt IDs[54] with the help of the external reference table available in the database. From the drug target list only those drugs that targeted human proteins were selected. From the original 6,718 drugs 3,926 such drugs were found, of which 3,626 had target proteins contained in our human interactome.

After comparison with the drug—side effect data from the SIDER database[2], we found that 597 drugs (with a total of 495 target proteins) had known side effects, while the remaining 3,029 drugs (with 1,231 target proteins) had no reported side effects to date.

**Protein and drug target data related to the two examined diseases: colorectal cancer and type 2 diabetes**

Genes involved in colorectal cancer were collected from the Cancer Gene Census[48] database, by selecting those proteins in the entire database that contained the word 'colorectal' in their 'Tumour Types' column. Genes related to type 2 diabetes were obtained from the article of Parchwani et al.[49]. The 18 genes involved in colorectal cancer and the 46 genes related to type 2 diabetes were then mapped to proteins marked by UniProt ID[54] with the help of the Protein Identifier Cross-Reference (PICR)[55] application. See Suppl. Tables 7 and 8 for the genes and their respective proteins involved in the two diseases. From these proteins, all 18 colorectal cancer-related but only 14 type 2 diabetes-related were contained in our interactome. Drugs used in treatment of colorectal cancer and diabetes and their drug targets were collected based on the drug indications in the DrugBank database[47]. See Suppl. Table 2 for the relevant keywords used. We found 11 drugs against colorectal cancer and 36 against type 2 diabetes, which all had valid targets. Drugs against colorectal cancer and type 2 diabetes had 33 and 42 target proteins, respectively, out of which 27 and 39, respectively, were contained in our human interactome.

**Other methods**

A number of Bash shell scripts were written to automate the network simulation experiments with Turbine. Statistical analysis of the results was performed with the R software package[56]. The Pajek software[57] was used to measure geodesic distances and centralities in the human interactome, the Cytoscape software[58] was used to create images of the human interactome and the Inkscape software[59] was used to create some other images.

## Acknowledgments

We thank members of the LINK-Group (www.linkgroup.hu) for helpful discussions. This work was supported by the Hungarian Scientific Research Fund [OTKA K83314]. T.K. was a grantee of the János Bolyai Scholarship of the Hungarian Academy of Sciences, and is supported by a fellowship in computational biology at The Genome Analysis Centre, in partnership with the Institute of Food Research, and strategically supported by BBSRC.

## Author contributions

P.C. initiated the project and conceived the research. A.R.P.L. performed all simulations and data analysis. D.T. and D.M. contributed in the assembly of databases. All (A.R.P.L., K.Z.S., D.T., D.M., K.L., T.K., P.C.) authors contributed to biological interpretation of the results. A.R.P.L. prepared the tables and figures. A.R.P.L. and P.C. wrote the manuscript text. All authors reviewed the manuscript.

**Competing financial interests:** The supporters had no role in study design, data collection and analysis, decision to publish, or preparation of the manuscript. The authors declare no competing financial interests.

## References

1. Fliri, A.F., Loging, W.T., Thadeio, P.F. & Volkmann, R.A. Analysis of drug-induced effect patterns to link structure and side effects of medicines. *Nat. Chem. Biol.* **1,** 389-397 (2005).
2. Kuhn, M., Campillos, M., Letunic, I., Jensen, L.J. & Bork, P. A side effect resource to capture phenotypic effects of drugs. *Mol. Syst. Biol.* **6**, 343 (2010).
3. Lounkine, E. *et al.* Large-scale prediction and testing of drug activity on side-effect targets. *Nature* **486,** 361–367 (2012).
4. Csermely, P., Korcsmáros, T., Kiss, H.J.M., London, G. & Nussinov, R. Structure and dynamics of molecular networks: A novel paradigm of drug discovery. *Pharmacol. Ther.* **138**, 333–408. (2013).
5. Yang, L., Luo, H., Chen, J., Xing, Q. & He, L. SePreSA: a server for the prediction of populations susceptible to serious adverse drug reactions implementing the methodology of a chemical-protein interactome. *Nucleic Acids Res.* **37**, W406–W412 (2009).
6. Yang, L., Xu, L. & He, L. A CitationRank algorithm inheriting Google technology designed to highlight genes responsible for serious adverse drug reaction. *Bioinformatics* **25**, 2244–2250 (2009).
7. Luo, H. *et al.* DRAR-CPI: a server for identifying drug repositioning potential and adverse drug reactions via the chemical-protein interactome. *Nucleic Acids Res.* **39**, W492–W498 (2011).
8. Oprea, T.I. *et al.* Associating drugs, targets and clinical outcomes into an integrated network affords a new platform for computer-aided drug repurposing. *Mol. Inform.* **30**, 100–111 (2011).
9. Lopes, P. *et al.* Gathering and exploring scientific knowledge in pharmacovigilance. *PLoS ONE* **8**, e83016 (2013).
10. Oliveira, J.L. *et al.* The EU-ADR Web Platform: delivering advanced pharmacovigilance tools. Pharmacoepidemiol. *Drug Saf.* **22**, 459–467 (2013).
11. Garten, Y., Tatonetti, N.P. & Altman, R.B. Improving the prediction of pharmacogenes using text-derived drug-gene relationships. *Pac. Symp. Biocomput.* 305–314 (2010).
12. Campillos, M., Kuhn, M., Gavin, A.C., Jensen, L.J. & Bork, P. Drug target identification using side-effect similarity. *Science* **321**, 263–266 (2008).
13. Yamanishi, Y., Kotera, M., Kanehisa, M., & Goto, S. Drug-target interaction prediction from chemical, genomic and pharmacological data in an integrated framework. *Bioinformatics* **26**, i246–i254 (2010).
14. Takarabe, M., Okuda, S., Itoh, M., Tokimatsu, T., Goto, S. & Kanehisa, M. Network analysis of adverse drug interactions. *Genome Inform.* **20**, 252–259 (2008).

15. Mizutani, S., Pauwels, E., Stoven, V., Goto, S. & Yamanishi, Y. Relating drug-protein interaction network with drug side effects. *Bioinformatics* **28,** i522–i528 (2012).
16. Iwata, H., Mizutani, S., Tabei, Y., Kotera, M., Goto, S. & Yamanishi Y. Inferring protein domains associated with drug side effects based on drug-target interaction network. *BMC Syst. Biol.* **7**, S18 (2013).
17. Lee, S., Lee, K.H., Song, M. & Lee, D. Building the process-drug-side effect network to discover the relationship between biological processes and side effects. *BMC Bioinformatics* **12**, S2 (2011).
18. Bauer-Mehren, A. *et al.* Automatic filtering and substantiation of drug safety signals. *PLoS Comput. Biol.* **8,** e1002457 (2012).
19. Schwartz, J.M. & Nacher, J.C. Local and global modes of drug action in biochemical networks. *BMC Chem. Biol.* **9**, 4 (2009).
20. Brouwers, L., Iskar, M., Zeller, G., van Noort, V. & Bork, P. Network neighbors of drug targets contribute to drug side-effect similarity. *PLoS ONE* **6**, e22187 (2011).
21. Nussinov, R., Tsai, C.-J. & Csermely, P. Allo-network drugs: harnessing allostery in cellular networks. *Trends Pharmacol. Sci,* **32**, 686–693 (2011).
22. Wang, J., Li, Z.-X., Qiu, C-X., Wang, D. & Cui, Q-H. The relationship between rational drug design and drug side effects. *Brief. Bioinform.* **13**, 377–382 (2012).
23. Nacher, J.C. & Schwartz, J.M. Modularity in protein complex and drug interactions reveals new polypharmacological properties. *PLoS ONE* **7**, e30028 (2012).
24. Hu, H., Myers, S., Colizza, V. & Vespignani A. WiFi networks and malware epidemiology. *Proc. Natl. Acad. Sci. USA* **106**, 1318–1323 (2009).
25. Wang, P., González, M.C., Hidalgo, C.A. & Barabási, A.L. Understanding the spreading patterns of mobile phone viruses. *Science* **324**, 1071–1076 (2009).
26. Brockmann, D. & Helbing, D. The hidden geometry of complex, network-driven contagion phenomena. *Science* **342**, 1337–1342 (2013).
27. Zanette, D.H. Critical behavior of propagation on small-world networks. *Phys. Rev. E* 64, 050901 (2001).
28. Valente, T.W. Network interventions. *Science* **337**, 49–53 (2012).
29. Banerjee, A., Chandrasekhar, A.G., Duflo, E. & Jackson, M.O. The diffusion of microfinance. *Science* **341**, 1236498 (2013).
30. Aral, S. & Walker, D. Identifying influential and susceptible members of social networks. *Science* **337**, 337–341 (2012).
31. Bray, D. & Duke, T. Conformational spread: the propagation of allosteric states in large multiprotein complexes. *Annu. Rev. Biophys. Biomol. Struct.* **33**, 53–73 (2004).
32. Antal, M.A., Böde, C. & Csermely, P. Perturbation waves in proteins and protein networks: applications of percolation and game theories in signaling and drug design. *Curr. Protein Pept. Sci.* **10**, 161–172 (2009).
33. Stojmirović, A., Bliskovsky, A. & Yu, Y.K. CytoITMprobe: a network information flow plugin for Cytoscape. *BMC Res. Notes* **5**, 237 (2012).
34. Cornelius, S.P., Kath, W.L. & Motter, A.E. Realistic control of network dynamics. *Nat. Commun.* **4,** 1942 (2013).
35. Szalay, K.Z. & Csermely, P. Perturbation centrality and Turbine: A novel centrality measure obtained using a versatile network dynamics tool. *PLoS ONE* **8**, e78059 (2013).
36. Szalay, K.Z., Nussinov, R. & Csermely, P. Attractor structures of signaling networks: Consequences of different conformational barcode dynamics and their relations to network-based drug design. *Mol. Info.* **33**, 463–468 (2014).

37. Jonsson, P.F. & Bates, P.A. Global topological features of cancer proteins in the human interactome. *Bioinformatics* **22,** 2291–2297 (2006).
38. Chuang, H.Y., Lee, E., Liu, Y.T., Lee, D. & Ideker, T. Network-based classification of breast cancer metastasis. *Mol. Syst. Biol.* **3,** 140 (2007).
39. Hase, T., Tanaka, H., Suzuki, Y., Nakagawa, S. & Kitano, H. Structure of protein interaction networks and their implications on drug design. *PLoS Comput. Biol.* **5**, e1000550 (2009).
40. Taylor, I.W., *et al.* Dynamic modularity in protein interaction networks predicts breast cancer outcome. *Nature Biotechn.* **27**, 199–204 (2009).
41. Sharma, A., Chavali, S., Tabassum, R., Tandon, N. & Bharadwaj, D. Gene prioritization in type 2 diabetes using domain interactions and network analysis. *BMC Genomics* **11**, 84 (2010).
42. Sun, J. & Zhao, Z. A comparative study of cancer proteins in the human protein-protein interaction network. *BMC Genomics* **11**, S5 (2010).
43. Rosado, J.O., Henriques, J.P., & Bonatto, D. A systems pharmacology analysis of major chemotherapy combination regimens used in gastric cancer treatment: predicting potential new protein targets and drugs. *Curr. Cancer Drug Targets* **11**, 849–869 (2011).
44. Xia, J., Sun, J., Jia, P. & Zhao, Z. Do cancer proteins really interact strongly in the human protein-protein interaction network? *Comput. Biol. Chem.* **35**, 121–125 (2011).
45. Serra-Musach, J. *et al.* Cancer develops, progresses and responds to therapies through restricted perturbation of the protein-protein interaction network. *Integr. Biol.* **4,** 1038–1048 (2012).
46. Franceschini, A. *et al.* STRING v9.1: protein-protein interaction networks, with increased coverage and integration. *Nucleic Acids Res.* **41**, D808–D815 (2012).
47. Knox, C. *et al.* DrugBank 3.0: a comprehensive resource for "omics" research on drugs. *Nucleic Acids Res.* **39**, D1035–D1041 (2011).
48. Forbes, S.A. *et al.* COSMIC: mining complete cancer genomes in the Catalogue of Somatic Mutations in Cancer. *Nucleic Acids Res.* **39**, D945–D950 (2010).
49. Parchwani, D., Murthy, S., Upadhyah, A. & Patel, D. Genetic factors in the etiology of type 2 diabetes: linkage analyses, candidate gene association, and genome-wide association – still a long way to go! *Natl. J. Physiol. Pharm. Pharmacol.* **3**, 57–68 (2013).
50. Yildirim, M.A., Goh, K.-I., Cusick, M.E., Barabási, A.-L. & Vidal, M. Drug-target network. *Nat. Biotechnol.* **25**, 1119–1126 (2007).
51. Mihalik, Á. & Csermely, P. Heat shock partially dissociates the overlapping modules of the yeast protein-protein interaction network: a systems level model of adaptation. *PLoS Comput. Biol.* **7**, e1002187 (2011).
52. Fazekas, D. *et al.* SignaLink 2 – A signaling pathway resource with multi-layered regulatory networks. *BMC Systems Biology* **7**, 7 (2013).
53. Veres, D. *et al.* ComPPI: a cellular compartment-specific database for protein-protein interaction network analysis. *Nucleic Acids Res.* **43**, D485-D493 (2015).
54. The UniProt Consortium. Reorganizing the protein space at the Universal Protein Resource (UniProt). *Nucleic Acids Res.* **40**, D71–D75 (2012).
55. Wein, S.P. *et al.* Improvements in the Protein Identifier Cross-Reference service. *Nucleic Acids Res.* **40**, W276–W280 (2012).
56. R Core Team. R: A language and environment for statistical computing. Vienna, Austria: R Foundation for Statistical Computing. Available: http://www.R-project.org/ (2013).

57. Bagatelj, V. & Mrvar, A. Pajek - Analysis and Visualization of Large Networks. in *Graph drawing software. Mathematics and visualization*. (eds Jünger, M. & Mutzel, P.) 77–103 (Springer, Berlin, 2003).
58. Shannon, P. et al. Cytoscape: a software environment for integrated models of biomolecular interaction networks. Genome Res. 13, 2498–2504 (2003).
59. The Inkscape Team. Inkscape. http://inkscape.org (2014).

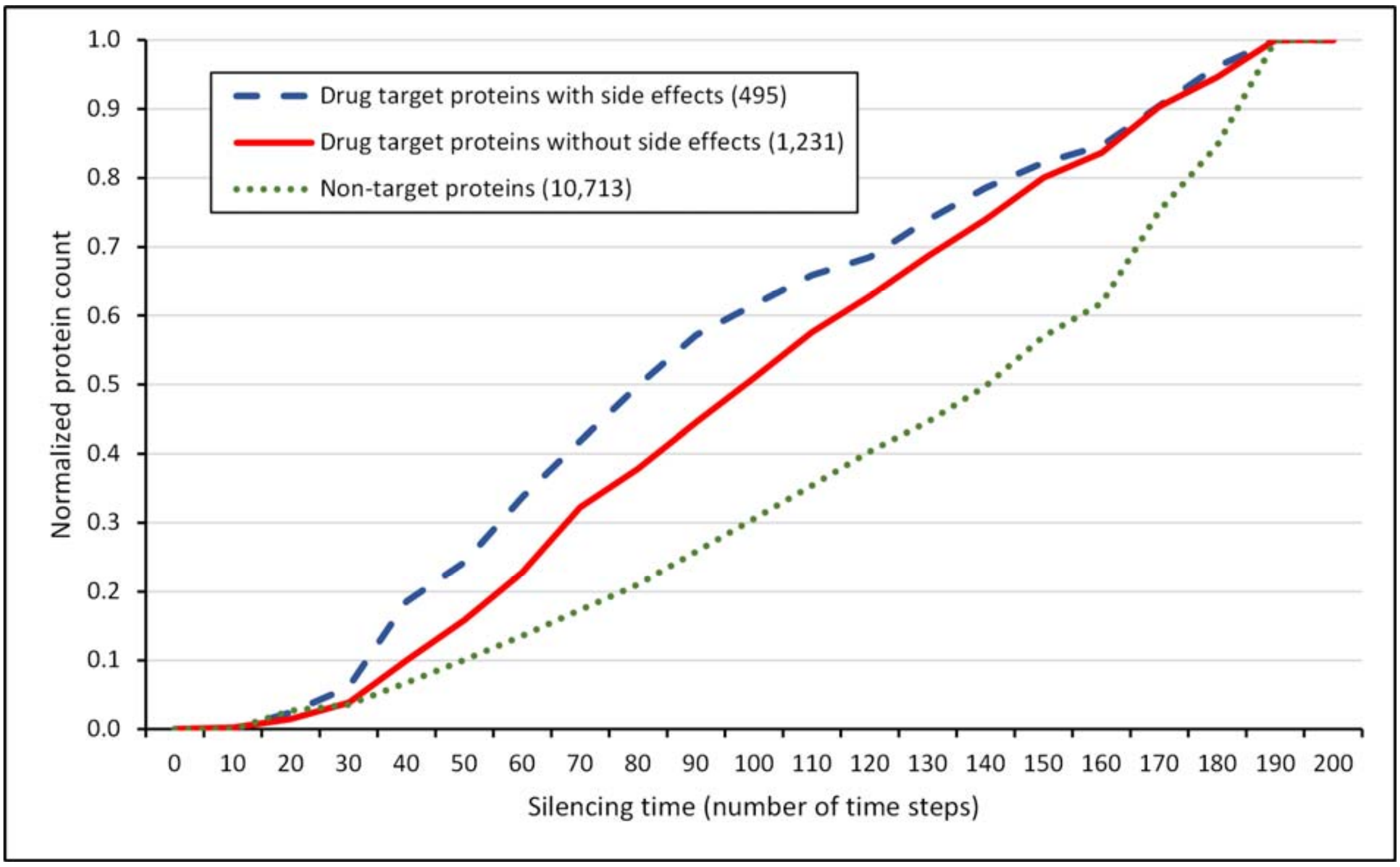

**Figure 1 │Cumulative silencing time distribution of drug targets and non-target proteins.** The diagram shows the cumulative distribution of the normalized number of proteins with given silencing times, which are drug targets with known side effects (blue dashed line), which are drug targets without known side effects (red solid line) and which are not drug targets (green dotted line). The number of proteins was normalized by dividing the number of proteins in each silencing time range by the total number of proteins allowing a better comparison. The total number of drug targets with and without side effects and non-target proteins was 495, 1,231 and 10,713, respectively. The human interactome containing 12,439 proteins and 174,666 edges was built from the STRING database[46], 1,726 human drug targets were obtained from the DrugBank database[47] and 99,423 drug-side effect pairs were taken from the SIDER database[2]. Silencing times were calculated separately for every protein/drug target with the Turbine program[35] as described in the Methods section using a starting energy of 1,000 and a dissipation value of 5 units. Statistical analysis was performed using the Mann-Whitney (Wilcoxon rank sum) test function of the R package[56]. There was a statistically significant difference ($p=1.677e-5$) between the silencing times of drug targets with known side effects and the silencing times of drug targets without reported side effects. The difference between the silencing times of drug targets and non-target proteins was also statistically significant ($p=2.2e-16$).

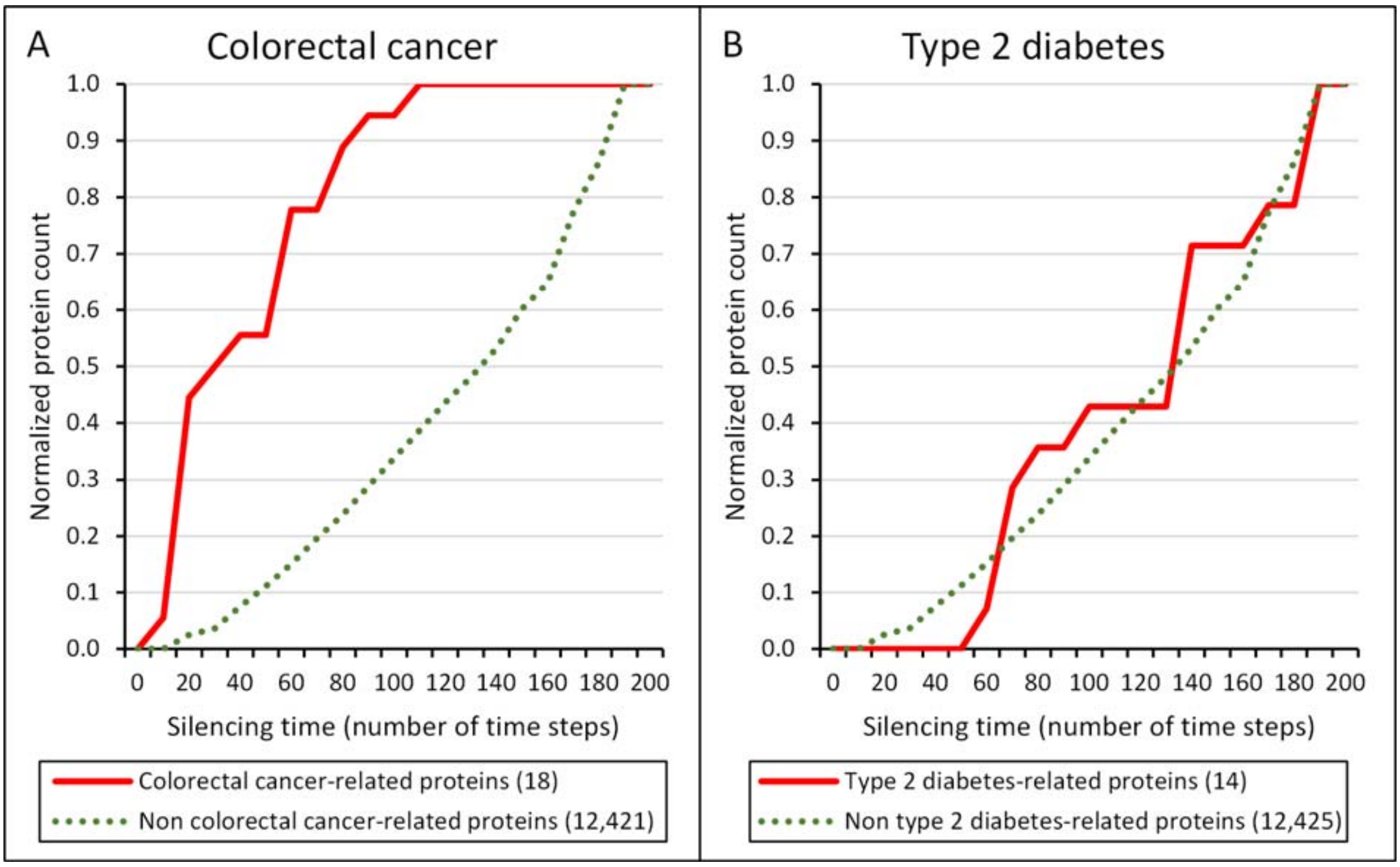

**Figure 2 │ Cumulative silencing time distribution of colorectal cancer- and type 2 diabetes mellitus-related proteins, as well as proteins, which are not related to these diseases.** The diagram shows the cumulative distribution of the normalized number of proteins with given silencing times, which are related to the disease (red line), as well as those, which are not related to the disease (green dotted line); for colorectal cancer (Panel A) and type 2 diabetes (Panel B). The number of proteins was normalized by dividing the number of proteins in each silencing time range by the total number of proteins allowing a better comparison. The total number of colorectal cancer-related proteins and type 2 diabetes-related proteins in the human interactome was 18 and 14, respectively. The human interactome containing 12,439 proteins and 174,666 edges was built from the STRING database[46]. Colorectal cancer- and type 2 diabetes-related proteins were obtained from the Cancer Gene Census database[48] and from the article of Parchwani et al.[49], respectively. Silencing times were calculated separately for every protein with the Turbine program[35] as described in the Methods section using a starting energy of 1,000 and a dissipation value of 5 units. Statistical analysis was performed using the Mann-Whitney (Wilcoxon rank sum) test function of the R package[56]. There was a statistically significant difference between the silencing times of disease-related and non-related proteins in case of colorectal cancer ($p=2.329e-9$) and but there was none in case of type 2 diabetes ($p=0.8343$).

**Table 1 | Average human interactome centralities of proteins related to colorectal cancer and type 2 diabetes**

| Centrality type | Disease-related proteins | | | Proteins, which are not related to any of the two diseases | | |
|---|---|---|---|---|---|---|
| | Colorectal cancer | Type 2 diabetes | Statistical difference between cancer- and diabetes-related proteins | Centrality value | Statistical difference from values of cancer-related proteins | Statistical difference from values of diabetes-related proteins |
| **Degree** (number of neighbours) | 159.5 | 9.000 | 7.09e-5 | 9.000 | 2.58e-9 | 0.830 |
| **Closeness centrality** (1/edge) | 0.357 | 0.294 | 3.46e-5 | 0.277 | 1.90e-10 | 0.122 |
| **Betweenness centrality** (fraction of shortest paths passing through the node) | 2.55e-3 | 1.16e-5 | 1.24e-4 | 1.34e-5 | 3.23e-9 | 0.922 |

The table shows the medians of the centralities of proteins related to colorectal cancer and type 2 diabetes (results were very similar, if instead of medians we used their arithmetic means; data not shown). The total number of colorectal cancer- and type 2 diabetes-related proteins was 18 and 14, respectively. Centrality values were calculated with the Pajek programme[57]. The human interactome containing 12,439 proteins and 174,666 edges was built from the STRING database[46]. Colorectal cancer-related proteins were obtained from the Cancer Gene Census database[48], type 2 diabetes-related proteins were obtained from the article of Parchwani et al.[49]. Statistical analysis was performed using the Wilcoxon rank sum (Mann-Whitney) test function of the R package[56].

**Table 2 | Average network distance of drug targets without and with known side effects used in the treatment of colorectal cancer and type 2 diabetes from the disease-associated proteins**

| **Protein group** | **Average network distance from disease-related proteins** (edges) |
|---|---|
| 24 drug targets without known side effects used in the treatment of colorectal cancer | 2.528 |
| 3 drug targets with known side effects used in the treatment of colorectal cancer | 2.389 |
| 14 drug targets without known side effects used in the treatment of type 2 diabetes | 3.250* |
| 25 drug targets with known side effects used in the treatment of type 2 diabetes | 3.234** |

*This value is significantly greater than the average network distance of drug targets without known side effects in colorectal cancer (p=1.062e-05). Statistical analysis was performed using the Welch (Student's) two sample t-test function of the R package[56].
**This value is significantly greater than the average network distance of drug targets with known side effects in colorectal cancer (p=0.005441). Statistical analysis was performed using the Welch (Student's) two sample t-test function of the R package[56].

The table shows the arithmetic mean of the average network distance between drug targets (with and without known side effects used in the treatment of colorectal cancer and type 2 diabetes) and the proteins related to the respective disease (results were very similar, if instead of arithmetic means we used the medians; data not shown). The total number of colorectal cancer- and diabetes-related proteins in the human interactome were 18 and 14, respectively. Average network distances were calculated as shortest paths using the Pajek programme[58]. Proteins were labelled by their UniProt ID[54]. Human interactome containing 12,439 proteins and 174,666 edges was built from the STRING database[46], 1,726 human drug targets were obtained from the DrugBank database[47] and 99,423 drug-side effect pairs were taken from the SIDER database[2]. Colorectal cancer- and type 2 diabetes-related proteins were obtained from the Cancer Gene Census database[48] and from the article of Parchwani et al.[49], respectively. We used the mean values and the t-test because of the near-normal distribution of the average network distances.

# Supplementary Information

## Targets of drugs are generally, and targets of drugs having side effects are specifically good spreaders of human interactome perturbations

**Áron R. Perez-Lopez[1,*], Kristóf Z. Szalay[1], Dénes Türei[2,+], Dezső Módos[2,3], Katalin Lenti[3], Tamás Korcsmáros[2,4,5] and Peter Csermely[1,#]**

**1** Department of Medical Chemistry, Semmelweis University, P.O. Box 260, H-1444 Budapest 8, Hungary; **2** Department of Genetics, Eötvös Loránd University, Pázmány P. s. 1C, H-1117 Budapest, Hungary; **3** Department of Morphology and Physiology, Faculty of Health Sciences, Semmelweis University, Vas u. 17, H-1088 Budapest, Hungary; **4** TGAC, The Genome Analysis Centre, Norwich, UK; **5** Gut Health and Food Safety Programme, Institute of Food Research, Norwich, UK

**Áron R. Perez Lopez is a high school student of the Apáczai Csere János High School of the Eötvös Loránd University, Papnövelde u. 4-6, H-1053 Budapest, Hungary; +current address: European Bioinformatics Institute (EMBL-EBI), Wellcome Trust Genome Campus, Cambridge CB10 1SD, UK*
*#Corresponding author, E-mail:* csermely.peter@med.semmelweis-univ.hu

## Table of Contents

## Supplementary Figures

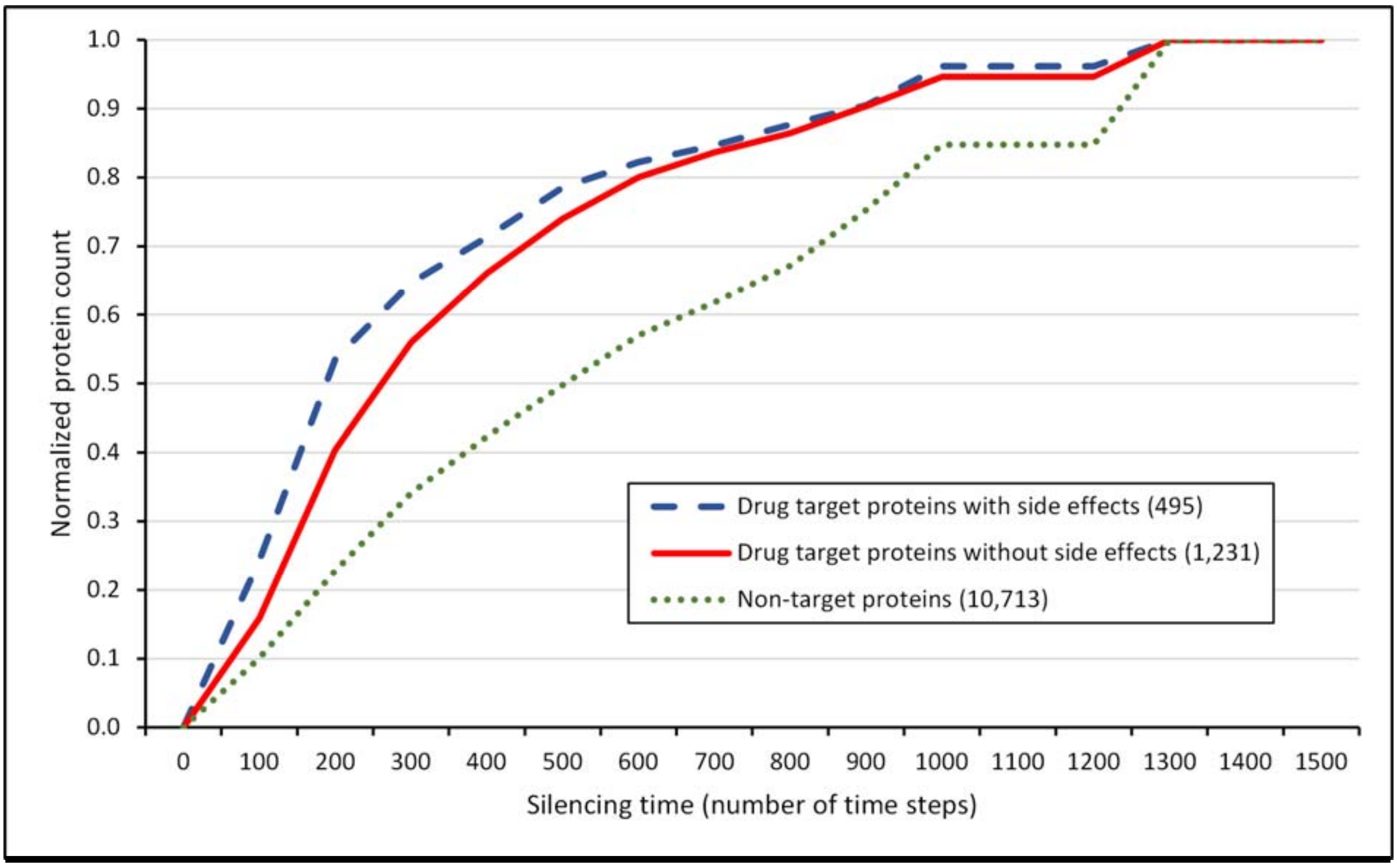

**Figure 1 | Cumulative silencing time distribution of drug targets and non-target proteins with a starting energy of 10,000 and a dissipation value of 5.** The diagram shows the cumulative distribution of the normalized number of proteins with given silencing times, which are drug targets with known side effects (blue dashed line), which are drug targets without known side effects (red solid line) and which are not drug targets (green dotted line). The number of proteins was normalized by dividing the number of proteins in each silencing time range by the total number of proteins allowing a better comparison. The total number of drug targets with and without side effects, and non-target proteins was 495, 1,231 and 10,713, respectively. The figure shows the 99.99% of all proteins (having a silencing time below 1500). The human interactome containing 12,439 proteins and 174,666 edges was built from the STRING database[1], 1,726 human drug targets were obtained from the DrugBank database[2] and 99,423 drug-side effect pairs were taken from the SIDER database[3]. Silencing times were calculated separately for every protein with the Turbine program[4] as described in the Methods section of the main text with a starting energy of 10,000 and a dissipation value of 5 units. Statistical analysis was performed using the Mann-Whitney (Wilcoxon rank sum) test function of the R package[5]. There was a statistically significant difference (p=1.701e-5) between the silencing times of drug targets with known side effects and the silencing times of drug targets without known side effects. The difference between the silencing times of drug targets and non-target proteins was also statistically significant (p=2.2e-16).

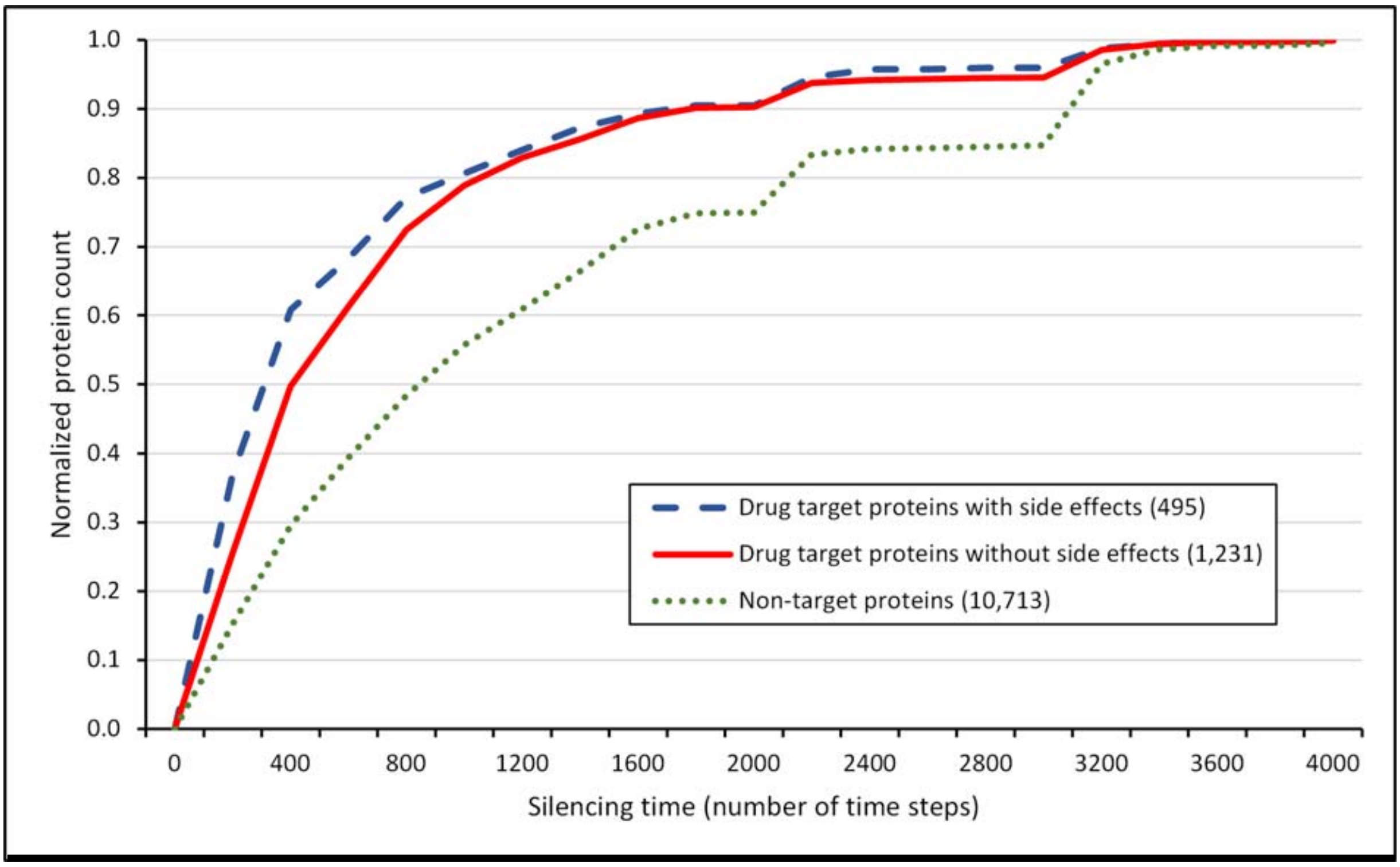

**Figure 2 | Cumulative silencing time distribution of drug targets and non-target proteins with a starting energy of 10,000 and a dissipation value of 1.** The diagram shows the cumulative distribution of the normalized number of proteins with given silencing times, which are drug targets with known side effects (blue dashed line), which are drug targets without known side effects (red solid line) and which are not drug targets (green dotted line). The number of proteins was normalized by dividing the number of proteins in each silencing time range by the total number of proteins allowing a better comparison. The total number of drug targets with and without side effects, and non-target proteins was 495, 1,231 and 10,713, respectively. The figure shows 99.61% of all proteins (having a silencing time below 4000). The human interactome containing 12,439 proteins and 174,666 edges was built from the STRING database[1], 1,726 human drug targets were obtained from the DrugBank database[2] and 99,423 drug-side effect pairs were taken from the SIDER database[3]. Silencing times were calculated separately for every protein with the Turbine program[4] as described in the Methods section of the main text with a starting energy of 10,000 and a dissipation value of 1 unit. Statistical analysis was performed using the Mann-Whitney (Wilcoxon rank sum) test function of the R package[5]. There was a statistically significant difference (p=9.635e-6) between the silencing times of drug targets with known side effects and the silencing times of drug targets without known side effects. The difference between the silencing times of drug targets and non-target proteins was also statistically significant (p=2.2e-16).

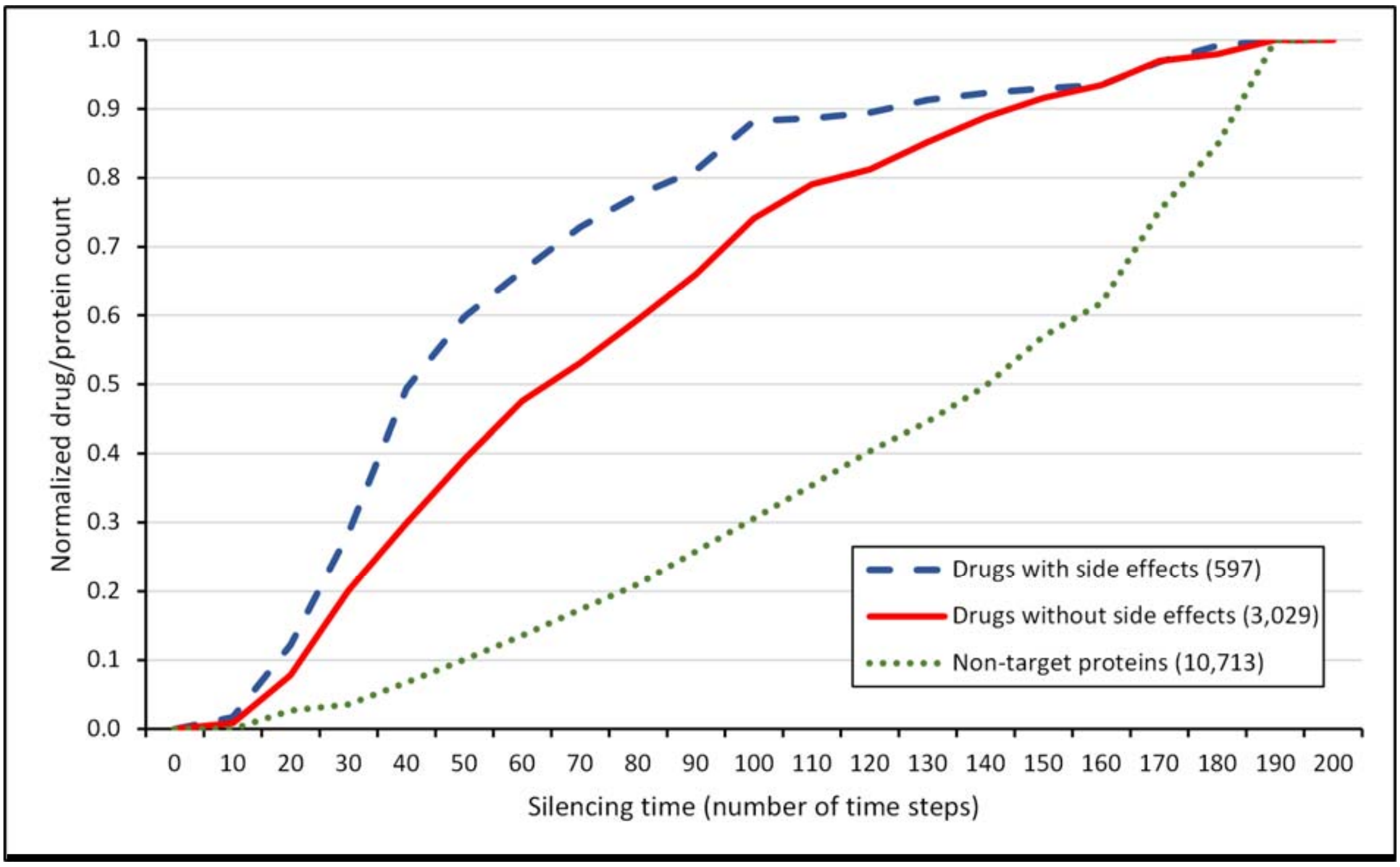

**Figure 3 │ Cumulative silencing time distribution of drugs and non-target proteins with starting energy of 1,000 and a dissipation value of 5 with distributed starting energy among multiple targets.** The diagram shows the cumulative silencing time distribution of the normalized number of drugs with known side effects (blue dashed line), drugs without known side effects (red solid line) and non-target proteins (green dotted line). The number of proteins/drugs was normalized by dividing the number of proteins/drugs in each silencing time range by the total number of proteins/drugs allowing a better comparison. The total number of drugs with and without side effects, and non-target proteins was 597, 3,029 and 10,713, respectively. The human interactome containing 12,439 proteins and 174,666 edges was built from the STRING database[1], 3,626 human drugs were obtained from the DrugBank database[2] and 99,423 drug-side effect pairs were taken from the SIDER database[3]. Silencing times were calculated separately for every protein/drug with the Turbine program[4] as described in the Methods section of the main text with a starting energy of 1000 and a dissipation value of 5 units. In case of drugs with multiple targets, the starting energy was distributed evenly among the drug targets. Statistical analysis was performed using the Mann-Whitney (Wilcoxon rank sum) test function of the R package[5]. There was a statistically significant difference (p=2.2e-16) between the silencing times of drugs with known side effects and the silencing times of drugs without known side effects. The difference between the silencing times of drugs and non-target proteins was also statistically significant (p=2.2e-16).

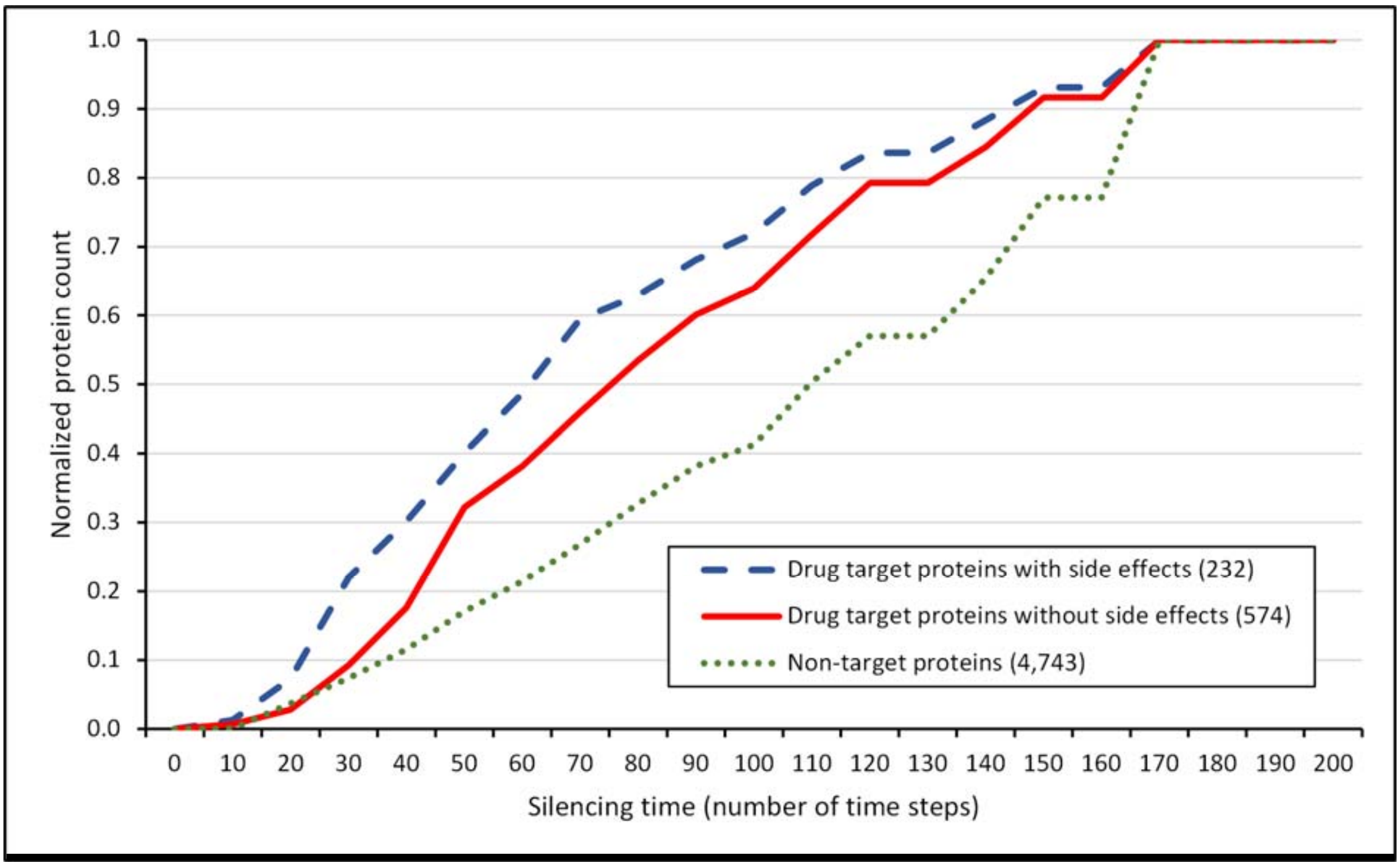

**Figure 4 | Cumulative silencing time distribution of drug target proteins and non-target proteins with a starting energy of 1000 and a dissipation value of 5 using a 50% smaller interactome.** The diagram shows the cumulative distribution of the normalized number of proteins with given silencing times, which are drug targets with known side effects (blue dashed line), which are drug targets without known side effects (red solid line) and which are not drug targets (green dotted line). The number of proteins was normalized by dividing the number of proteins in each silencing time range by the total number of proteins allowing a better comparison. The total number of drug targets with and without side effects, and non-target proteins was 495, 1,231 and 10,713, respectively. The human interactome containing 12,439 proteins and 174,666 edges was built from the STRING database[1], 1,726 human drug targets were obtained from the DrugBank database[2] and 99,423 drug-side effect pairs were taken from the SIDER database[3]. 50% of the original interactome proteins were deleted randomly. The giant component of the remaining interactome contained 5,549 proteins (45%), 806 drug target proteins total (47%) and 232 drug targets with known side effects (47%). Silencing times were calculated separately for every protein with the Turbine program[4] as described in the Methods section of the main text with a starting energy of 1,000 and a dissipation value of 5 units. Statistical analysis was performed using the Mann-Whitney (Wilcoxon rank sum) test function of the R package[5]. There was a statistically significant difference (p=3.368e-4) between the silencing times of drug targets with known side effects and the silencing times of drug targets without known side effects. The difference between the silencing times of drug targets and non-target proteins was also statistically significant (p=2.2e-16).

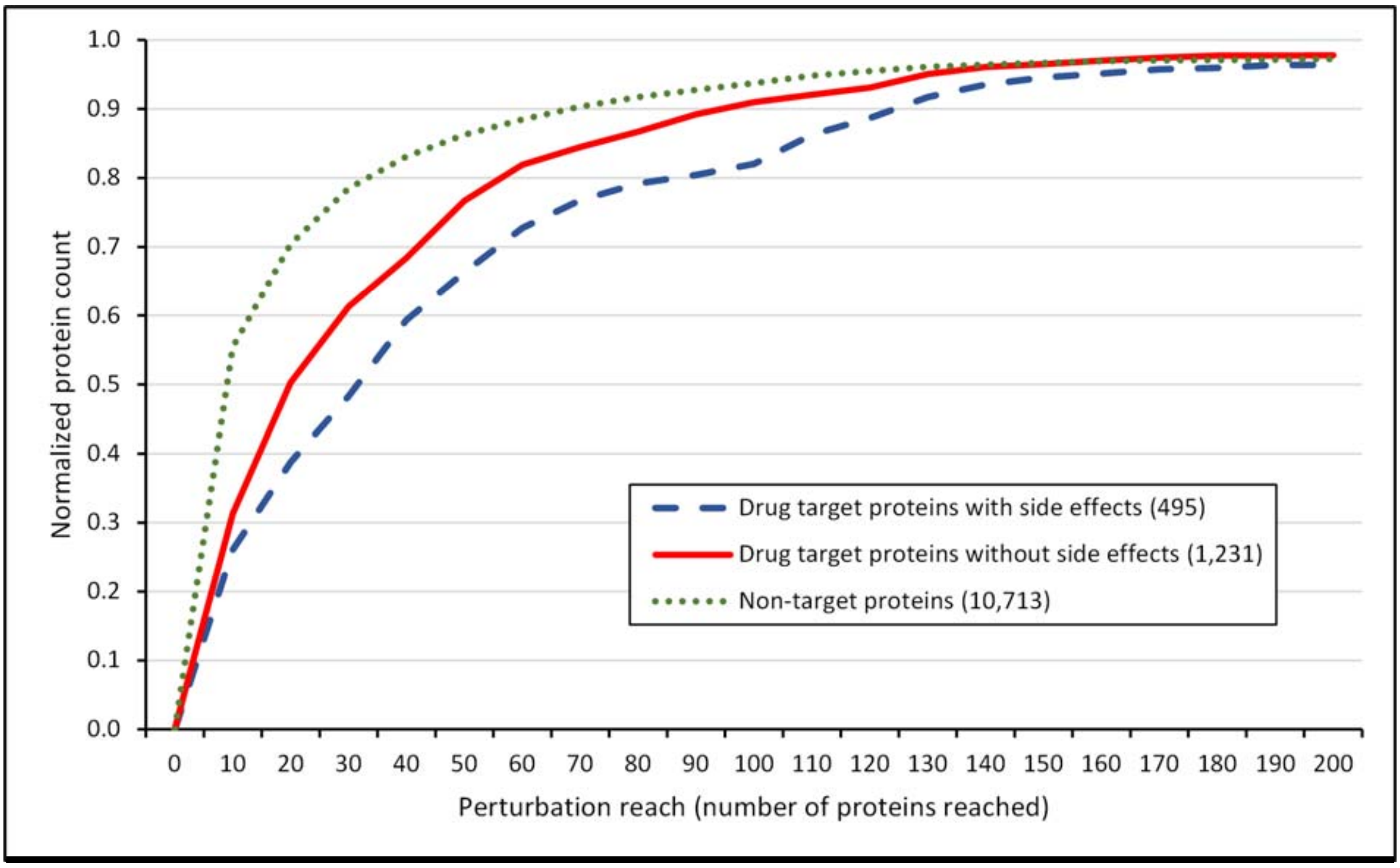

**Figure 5│Cumulative perturbation reach distribution of drug targets and non-target proteins with a starting energy of 10,000 and a dissipation value of 5.** The diagram shows the cumulative distribution of the normalized number of proteins with given perturbation reach values, which are drug targets with known side effects (blue dashed line), which are drug targets without known side effects (red solid line) and which are not drug targets (green dotted line). The number of proteins was normalized by dividing the number of proteins in each perturbation reach range by the total number of proteins allowing a better comparison. The total number of drug targets with and without side effects, and non-target proteins was 495, 1,231 and 10,713, respectively. The figure shows 97.25% of all proteins (having a perturbation reach below 200 proteins reached). The human interactome containing 12,439 proteins and 174,666 edges was built from the STRING database[1], 1,726 human drug targets were obtained from the DrugBank database[2] and 99,423 drug-side effect pairs were taken from the SIDER database[3]. Perturbation reach values were calculated separately for every protein with the Turbine program[4] as described in the Methods section of the main text with a starting energy of 10,000 and a dissipation value of 5 units. Statistical analysis was performed using the Mann-Whitney (Wilcoxon rank sum) test function of the R package[5]. There was a statistically significant difference (p=1.663e-5) between the perturbation reach values of drug targets with known side effects and the perturbation reach values of drug targets without known side effects. The difference between the perturbation reach values of drug targets and non-target proteins was also statistically significant (p=2.2e-16).

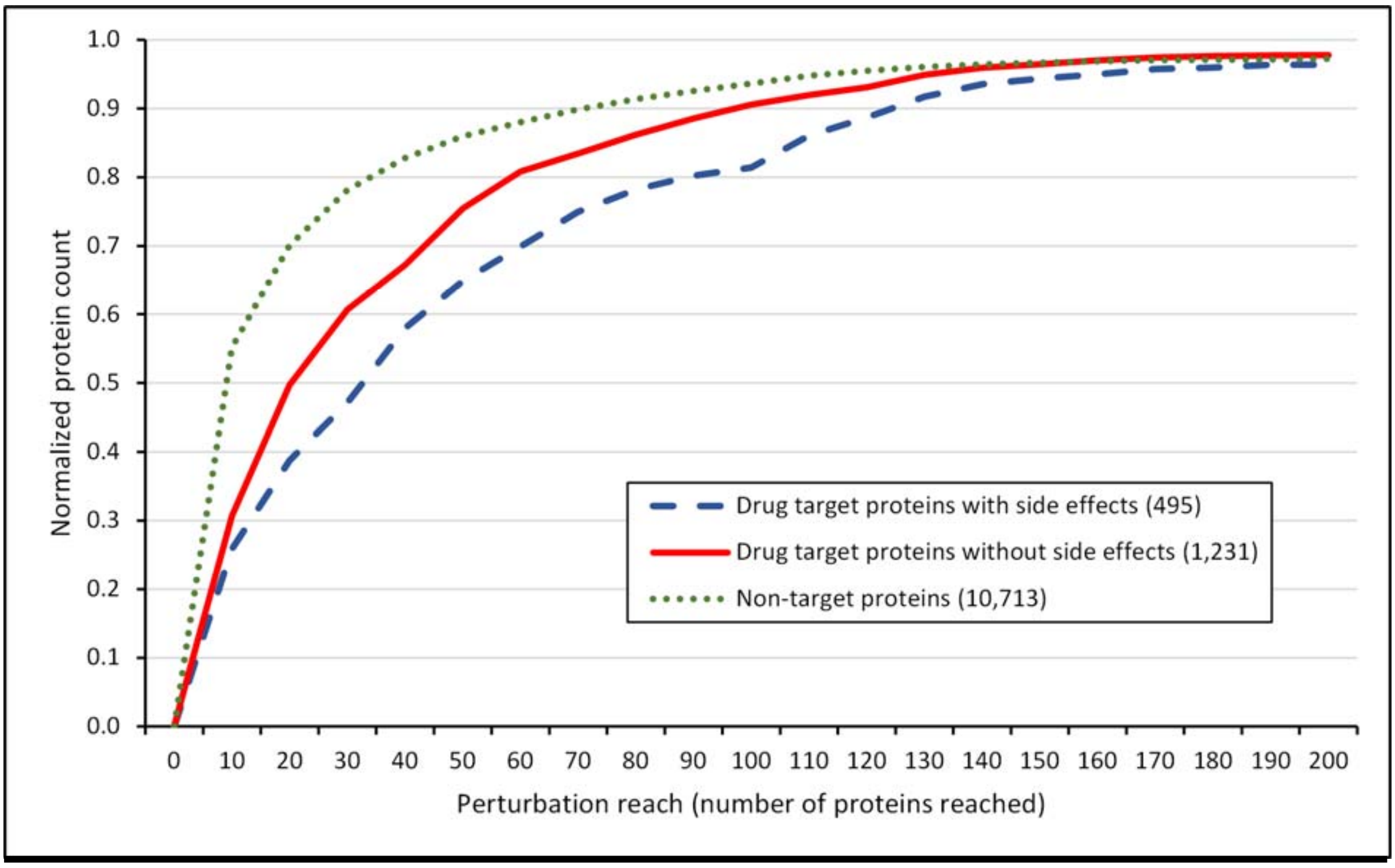

**Figure 6│Cumulative perturbation reach distribution of drug targets and non-target proteins with a starting energy of 10,000 and a dissipation value of 1.** The diagram shows the cumulative distribution of the normalized number of proteins with given perturbation reach values, which are drug targets with known side effects (blue dashed line), which are drug targets without known side effects (red solid line) and which are not drug targets (green dotted line). The number of proteins was normalized by dividing the number of proteins in each perturbation reach range by the total number of proteins allowing a better comparison. The total number of drug targets with and without side effects, and non-target proteins was 495, 1,231 and 10,713, respectively. The figure shows 97.25% of all proteins (having a perturbation reach below 200 proteins reached). The human interactome containing 12,439 proteins and 174,666 edges was built from the STRING database[1], 1,726 human drug targets were obtained from the DrugBank database[2] and 99,423 drug-side effect pairs were taken from the SIDER database[3]. Perturbation reach values were calculated separately for every protein with the Turbine program[4] as described in the Methods section of the main text with a starting energy of 10,000 and a dissipation value of 1 unit. Statistical analysis was performed using the Mann-Whitney (Wilcoxon rank sum) test function of the R package[5]. There was a statistically significant difference (p=1.49e-5) between the perturbation reach values of drug targets with known side effects and the perturbation reach values of drug targets without known side effects. The difference between the perturbation reach values of drug targets and non-target proteins was also statistically significant (p=2.2e-16).

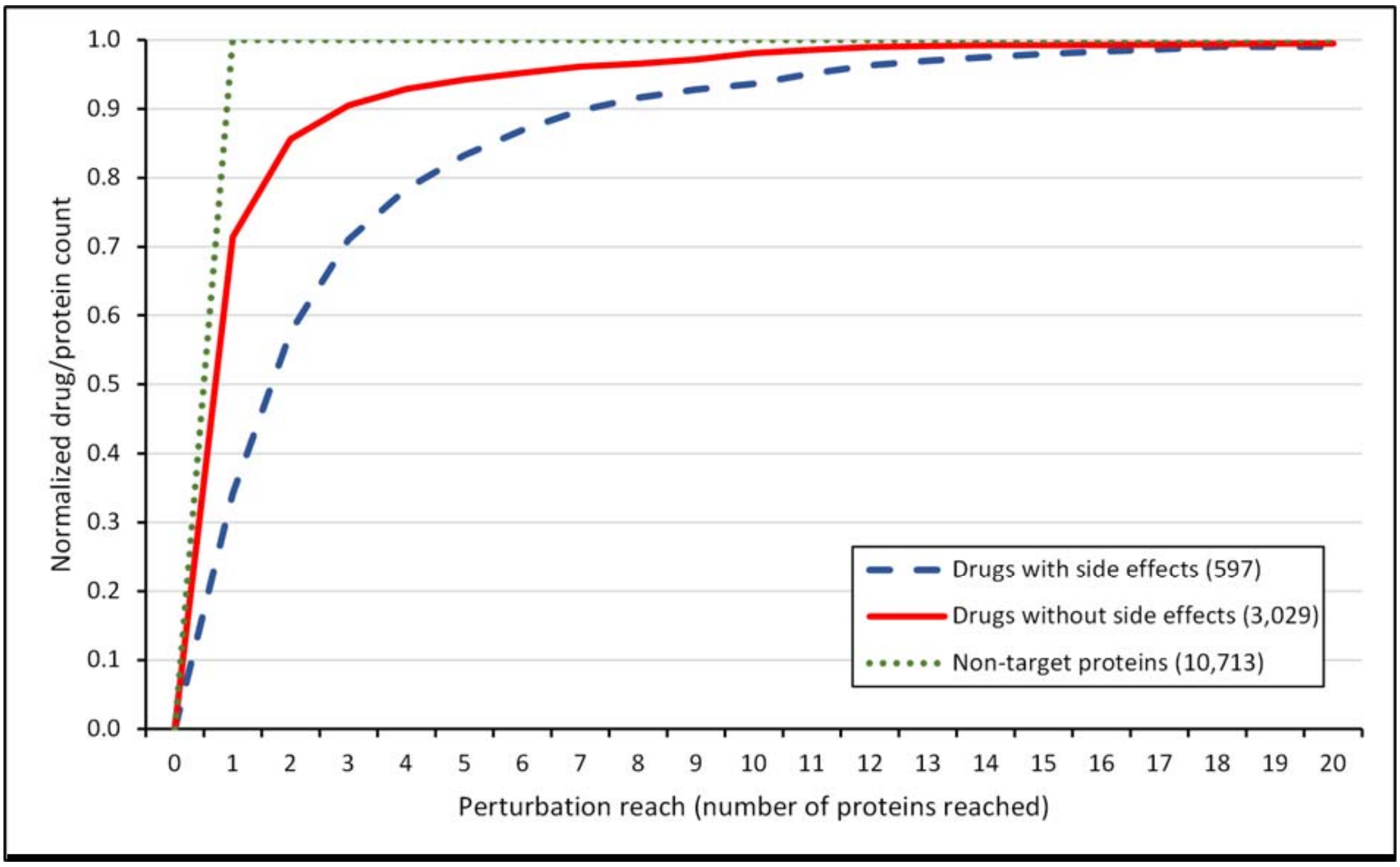

**Figure 7 | Cumulative perturbation reach distribution of drugs and non-target proteins with starting energy of 10,000 and a dissipation value of 1 with distributed starting energy among multiple targets.** The diagram shows the cumulative perturbation reach distribution of the normalized number of drugs with known side effects (blue dashed line), drugs without known side effects (red solid line) and non-target proteins (green dotted line). The number of proteins/drugs was normalized by dividing the number of proteins/drugs in each perturbation reach range by the total number of proteins/drugs allowing a better comparison. The total number of drugs with and without side effects, and non-target proteins was 597, 3,029 and 10,713, respectively. The figure shows 99.58% of all proteins/drugs (having a perturbation reach below 400 proteins reached). The human interactome containing 12,439 proteins and 174,666 edges was built from the STRING database[1], 3,626 human drugs were obtained from the DrugBank database[2] and 99,423 drug-side effect pairs were taken from the SIDER database[3]. Perturbation reach values were calculated separately for every protein/drug with the Turbine program[4] as described in the Methods section of the main text with a starting energy of 10,000 and a dissipation value of 1 unit. In case of drugs with multiple targets, the starting energy was distributed evenly among the drug targets. Statistical analysis was performed using the Mann-Whitney (Wilcoxon) test function of the R package[5]. There was a statistically significant difference (p=6.176e-8) between the perturbation reach values of drugs with known side effects and the perturbation reach values of drugs without known side effects. The difference between the perturbation reach values of drugs and non-target proteins was also statistically significant (p=2.2e-16).

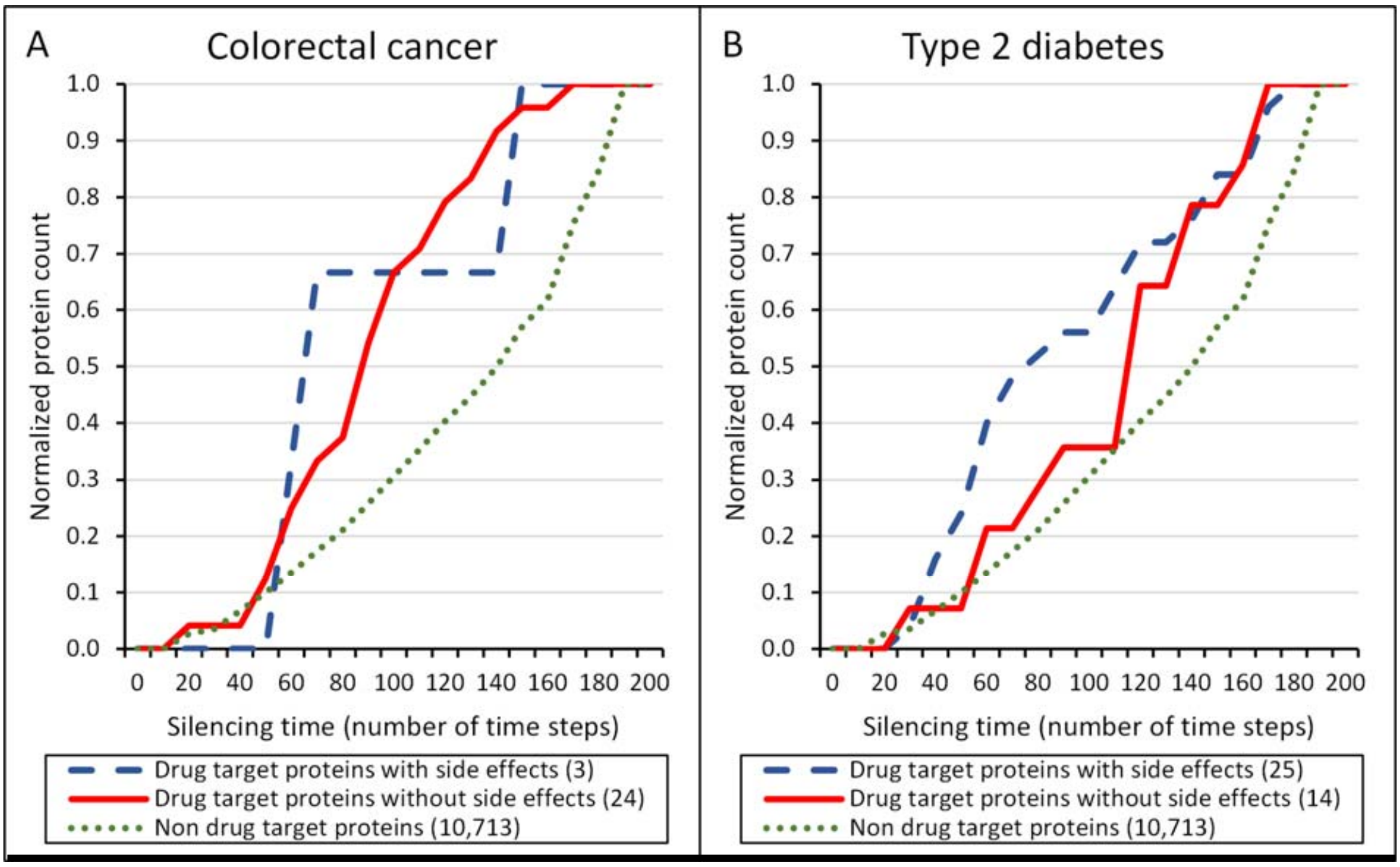

**Figure 8 | Cumulative silencing time distribution of targets of drugs used in the treatment of colorectal cancer and type 2 diabetes mellitus.** The diagram shows the cumulative distribution of the normalized number of proteins with given silencing times, which are drug targets used in the treatment of the disease with known side effects (blue dashed line), which are drug targets used in the treatment of the disease without known side effects (red solid line) and which are not drug targets (green dotted line); for colorectal cancer (Panel A) and type 2 diabetes (Panel B). The number of proteins was normalized by dividing the number of proteins in each silencing time range by the total number of proteins allowing a better comparison. The total number of drug targets used in the treatment of colorectal cancer with and without side effects was 3 and 24, respectively, while for type 2 diabetes the total number of drug targets was 25 and 14, respectively. The human interactome containing 12,439 proteins and 174,666 edges was built from the STRING database[1], 1,726 human drug targets were obtained from the DrugBank database[2] and 99,423 drug-side effect pairs were taken from the SIDER database[3]. Silencing times were calculated separately for every protein with the Turbine program[4] as described in the Methods section of the main text with a starting energy of 1,000 and a dissipation value of 5 units. Statistical analysis was performed using the Mann-Whitney-Wilcoxon test of the R package[5]. No statistically significant difference could be shown between silencing times of targets with known side effects and silencing times of targets without known side effects of drugs used in the treatment of colorectal cancer (p=1) and type 2 diabetes (p=0.2593). However, the difference between the silencing times of drug targets and non-target proteins was statistically significant for drug targets used in the treatment of both colorectal cancer (p=3.367e-5) and type 2 diabetes (p=5.88e-5).

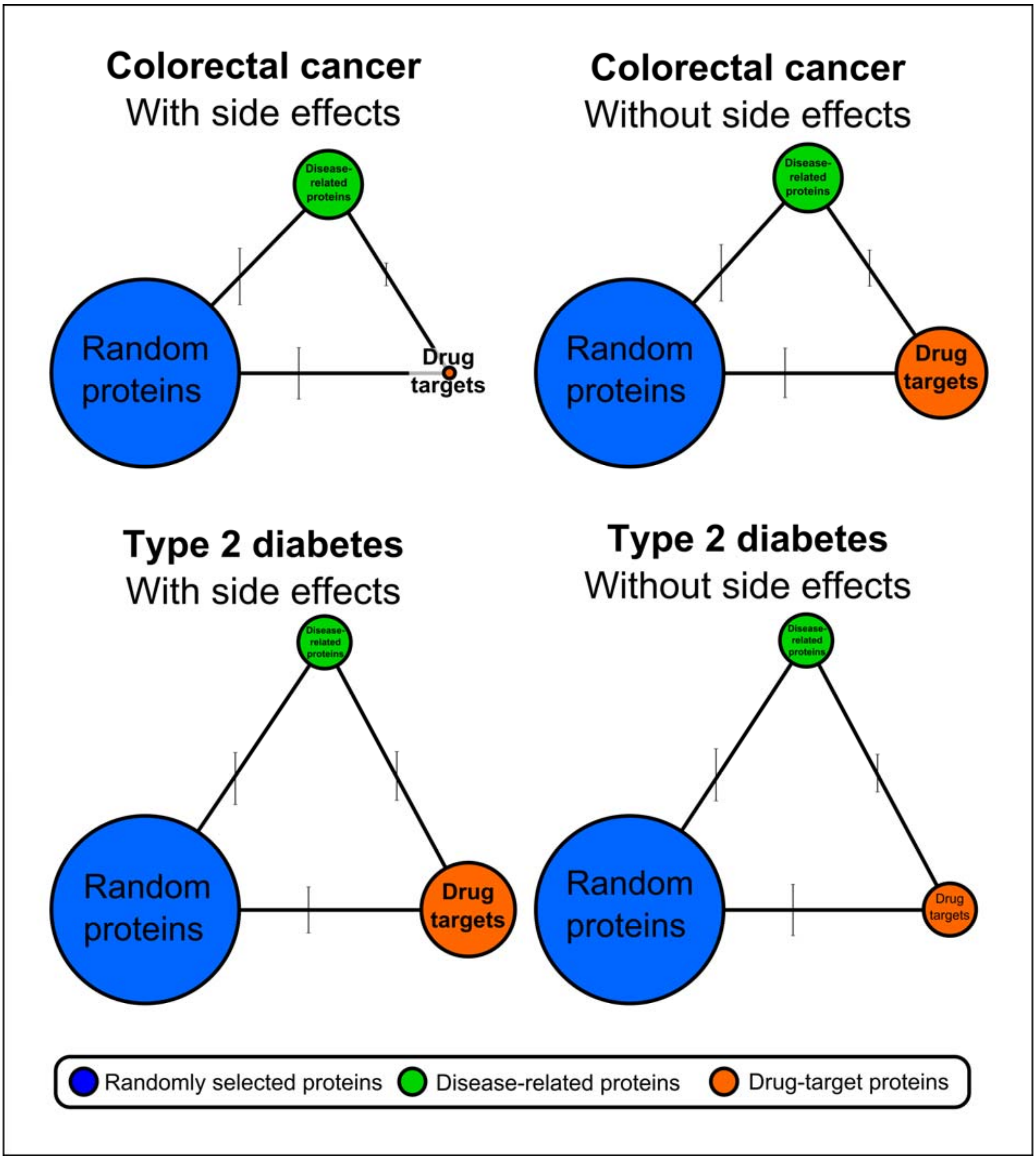

**Figure 9│Human interactome distance between drug targets used in the treatment of colorectal cancer and type 2 diabetes, between proteins related to these diseases and randomly selected proteins.** The figure shows the average human interactome distances between the following proteins: drug targets used in the treatment of colorectal cancer and type 2 diabetes with and without side effects (orange circles), proteins related to these diseases (green circles) and randomly selected proteins (blue circles). The sides of the triangles (the distance between the centres of the circles) are proportional to the average number of human interactome

edges between the respective protein groups, while the vertical lines associated with the sides of the triangles correspond to the standard deviation (SD). The average distance between randomly selected proteins and disease-related proteins was 2.82 edges (SD: 0.601) for colorectal cancer and 3.43 edges (SD: 0.557) for type 2 diabetes; between randomly selected proteins and drug targets with side effects was 3.24 edges (SD: 0.551) for colorectal cancer and 3.44 edges (SD: 0.490) for type 2 diabetes; between randomly selected proteins and drug targets without side effects was 3.32 edges (SD: 0.533) for colorectal cancer and 3.41 edges (SD: 0.545) for type 2 diabetes; between disease-related proteins and drug targets with side effects was 2.39 edges (SD: 0.242) for colorectal cancer and 3.23 edges (SD: 0.522) for type 2 diabetes; between disease-related proteins and drug targets without side effects was 2.53 edges (SD: 0.388) for colorectal cancer and 3.25 edges (SD: 0.402) for type 2 diabetes. Sizes of the circles are proportional to the number of proteins contained in each group. There were 50 randomly selected proteins; 18 colorectal cancer-related and 14 type 2 diabetes-related proteins; 3 drug targets with and 24 drug targets without side effects used in the treatment of colorectal cancer; 25 drug targets with and 14 drug targets without side effects used in the treatment of type 2 diabetes. The human interactome containing 12,439 proteins and 174,666 edges was built from the STRING database[1], 1,726 human drug targets were obtained from the DrugBank database[2] and 99,423 drug-side effect pairs were taken from the SIDER database[3]. Network distances were calculated as shortest paths using the Pajek programme[6] as described in the Methods section of the main text and are detailed in Tables 10-13. The figure was created using Inkscape[7].

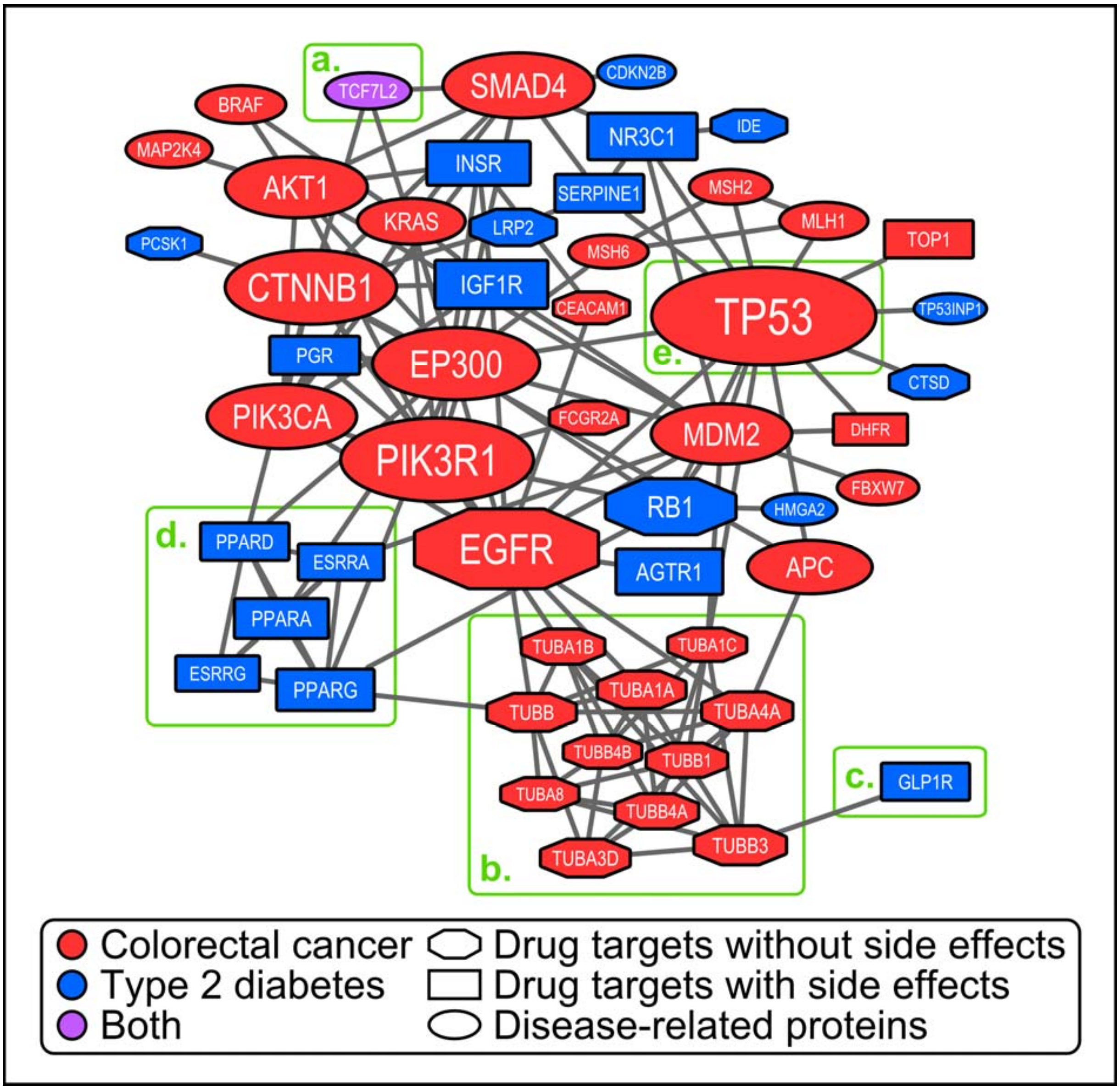

**Figure 10 | Human protein-protein interaction network of the proteins related to colorectal cancer and type 2 diabetes and the drug targets used in the treatment of these diseases.** The figure shows the giant component of the human protein-protein interaction network containing the proteins related to colorectal cancer and type 2 diabetes mellitus and the drug targets used in the treatment of these diseases. Red nodes represent proteins or drug targets related to colorectal cancer, blue nodes represent those related to type 2 diabetes, while purple nodes represent those related to both. Ellipses, octagons and squares represent proteins related to diseases, drug targets without known side effects and drug targets with known side effects, respectively. Node highlighted by green box (a.) is the TCF7L2 protein related to both diseases, which is the transcription factor 7-like 2 participating in the Wnt signalling pathway and modulating MYC expression. The highly interconnected node cluster highlighted by green box (b.) contains 11 drug targets without known side effects used in the treatment of colorectal cancer, which are all

tubuline chain proteins. Node highlighted by green box (c.) representing protein GLP1R, the glucagon-like peptide 1 receptor, is connected only to node TUBB3 of the tubuline cluster (b.). The highly interconnected node cluster highlighted by green box (d.) contains 5 drug targets with known side effects used in the treatment of type 2 diabetes which are the peroxisome proliferator-activated receptors alpha (PPARA), gamma (PPARG) and delta (PPARD) and the estrogen-related receptors alpha (ESRRA) and gamma (ESSRG). The network hub highlighted by green box (e.) is TP53, the cellular tumour antigen p53. Node sizes are proportional to the degrees of the respective proteins in the full human protein-protein interaction network. All proteins here are referenced by their UniProt ID[9]. The human interactome containing 12,439 proteins and 174,666 edges was built from the STRING database[1], 1,726 human drug targets were obtained from the DrugBank database[2] and 99,423 drug-side effect pairs were taken from the SIDER database[3]. Node degrees were calculated with the Pajek programme[6] as described in the Methods section of the main text. The figure was created using Cytoscape[8] and Inkscape[7].

## Supplementary Tables

### Table 1 │ Drugs obtained from the DrugBank database, which have known side effects in the SIDER database

| DBID | Drug Name | DBID | Drug Name | DBID | Drug Name |
|---|---|---|---|---|---|
| ***DB00001*** | Lepirudin | ***DB00210*** | Adapalene | ***DB00289*** | Atomoxetine |
| ***DB00006*** | Bivalirudin | ***DB00211*** | Midodrine | ***DB00292*** | Etomidate |
| ***DB00046*** | Insulin Lispro | ***DB00213*** | Pantoprazole | ***DB00293*** | Raltitrexed |
| ***DB00047*** | Insulin Glargine | ***DB00214*** | Torasemide | ***DB00295*** | Morphine |
| ***DB00050*** | Cetrorelix | ***DB00215*** | Citalopram | ***DB00296*** | Ropivacaine |
| ***DB00063*** | Eptifibatide | ***DB00216*** | Eletriptan | ***DB00297*** | Bupivacaine |
| ***DB00106*** | Abarelix | ***DB00218*** | Moxifloxacin | ***DB00302*** | Tranexamic Acid |
| ***DB00115*** | Cyanocobalamin | ***DB00222*** | Glimepiride | ***DB00307*** | Bexarotene |
| ***DB00125*** | L-Arginine | ***DB00227*** | Lovastatin | ***DB00308*** | Ibutilide |
| ***DB00152*** | Thiamine | ***DB00228*** | Enflurane | ***DB00310*** | Chlorthalidone |
| ***DB00162*** | Vitamin A | ***DB00231*** | Temazepam | ***DB00312*** | Pentobarbital |
| ***DB00175*** | Pravastatin | ***DB00240*** | Alclometasone | ***DB00313*** | Valproic Acid |
| ***DB00176*** | Fluvoxamine | ***DB00242*** | Cladribine | ***DB00315*** | Zolmitriptan |
| ***DB00177*** | Valsartan | ***DB00243*** | Ranolazine | ***DB00316*** | Acetaminophen |
| ***DB00178*** | Ramipril | ***DB00246*** | Ziprasidone | ***DB00317*** | Gefitinib |
| ***DB00180*** | Flunisolide | ***DB00247*** | Methysergide | ***DB00318*** | Codeine |
| ***DB00182*** | Amphetamine | ***DB00248*** | Cabergoline | ***DB00320*** | Dihydroergotamine |
| ***DB00184*** | Nicotine | ***DB00252*** | Phenytoin | ***DB00321*** | Amitriptyline |
| ***DB00185*** | Cevimeline | ***DB00253*** | Medrysone | ***DB00323*** | Tolcapone |
| ***DB00186*** | Lorazepam | ***DB00257*** | Clotrimazole | ***DB00324*** | Fluorometholone |
| ***DB00187*** | Esmolol | ***DB00264*** | Metoprolol | ***DB00327*** | Hydromorphone |
| ***DB00188*** | Bortezomib | ***DB00268*** | Ropinirole | ***DB00328*** | Indomethacin |
| ***DB00191*** | Phentermine | ***DB00273*** | Topiramate | ***DB00331*** | Metformin |
| ***DB00193*** | Tramadol | ***DB00276*** | Amsacrine | ***DB00332*** | Ipratropium bromide |
| ***DB00195*** | Betaxolol | ***DB00277*** | Theophylline | ***DB00333*** | Methadone |
| ***DB00197*** | Troglitazone | ***DB00278*** | Argatroban | ***DB00334*** | Olanzapine |
| ***DB00198*** | Oseltamivir | ***DB00280*** | Disopyramide | ***DB00335*** | Atenolol |
| ***DB00200*** | Hydroxocobalamin | ***DB00281*** | Lidocaine | ***DB00337*** | Pimecrolimus |
| ***DB00201*** | Caffeine | ***DB00282*** | Pamidronate | ***DB00338*** | Omeprazole |
| ***DB00202*** | Succinylcholine | ***DB00284*** | Acarbose | ***DB00343*** | Diltiazem |
| ***DB00204*** | Dofetilide | ***DB00285*** | Venlafaxine | ***DB00344*** | Protriptyline |
| ***DB00205*** | Pyrimethamine | ***DB00286*** | Conjugated Estrogens | ***DB00346*** | Alfuzosin |
| ***DB00206*** | Reserpine | ***DB00287*** | Travoprost | ***DB00349*** | Clobazam |
| ***DB00208*** | Ticlopidine | ***DB00288*** | Amcinonide | ***DB00350*** | Minoxidil |

| DBID | Drug Name |
|---|---|
| ***DB00351*** | Megestrol |
| ***DB00356*** | Chlorzoxazone |
| ***DB00357*** | Aminoglutethimide |
| ***DB00358*** | Mefloquine |
| ***DB00360*** | Tetrahydrobiopterin |
| ***DB00361*** | Vinorelbine |
| ***DB00363*** | Clozapine |
| ***DB00364*** | Sucralfate |
| ***DB00367*** | Levonorgestrel |
| ***DB00368*** | Norepinephrine |
| ***DB00370*** | Mirtazapine |
| ***DB00371*** | Meprobamate |
| ***DB00373*** | Timolol |
| ***DB00374*** | Treprostinil |
| ***DB00376*** | Trihexyphenidyl |
| ***DB00377*** | Palonosetron |
| ***DB00379*** | Mexiletine |
| ***DB00380*** | Dexrazoxane |
| ***DB00381*** | Amlodipine |
| ***DB00382*** | Tacrine |
| ***DB00384*** | Triamterene |
| ***DB00388*** | Phenylephrine |
| ***DB00390*** | Digoxin |
| ***DB00393*** | Nimodipine |
| ***DB00396*** | Progesterone |
| ***DB00398*** | Sorafenib |
| ***DB00401*** | Nisoldipine |
| ***DB00404*** | Alprazolam |
| ***DB00408*** | Loxapine |
| ***DB00411*** | Carbachol |
| ***DB00412*** | Rosiglitazone |
| ***DB00413*** | Pramipexole |
| ***DB00418*** | Secobarbital |
| ***DB00419*** | Miglustat |
| ***DB00421*** | Spironolactone |
| ***DB00422*** | Methylphenidate |
| ***DB00423*** | Methocarbamol |
| ***DB00425*** | Zolpidem |

| DBID | Drug Name |
|---|---|
| ***DB00431*** | Lindane |
| ***DB00433*** | Prochlorperazine |
| ***DB00434*** | Cyproheptadine |
| ***DB00437*** | Allopurinol |
| ***DB00439*** | Cerivastatin |
| ***DB00440*** | Trimethoprim |
| ***DB00441*** | Gemcitabine |
| ***DB00444*** | Teniposide |
| ***DB00446*** | Chloramphenicol |
| ***DB00448*** | Lansoprazole |
| ***DB00449*** | Dipivefrin |
| ***DB00450*** | Droperidol |
| ***DB00454*** | Meperidine |
| ***DB00457*** | Prazosin |
| ***DB00458*** | Imipramine |
| ***DB00459*** | Acitretin |
| ***DB00461*** | Nabumetone |
| ***DB00462*** | Methylscopolamine |
| ***DB00465*** | Ketorolac |
| ***DB00471*** | Montelukast |
| ***DB00472*** | Fluoxetine |
| ***DB00474*** | Methohexital |
| ***DB00475*** | Chlordiazepoxide |
| ***DB00476*** | Duloxetine |
| ***DB00477*** | Chlorpromazine |
| ***DB00480*** | Lenalidomide |
| ***DB00481*** | Raloxifene |
| ***DB00482*** | Celecoxib |
| ***DB00484*** | Brimonidine |
| ***DB00486*** | Nabilone |
| ***DB00489*** | Sotalol |
| ***DB00490*** | Buspirone |
| ***DB00491*** | Miglitol |
| ***DB00492*** | Fosinopril |
| ***DB00494*** | Entacapone |
| ***DB00496*** | Darifenacin |
| ***DB00497*** | Oxycodone |
| ***DB00499*** | Flutamide |

| DBID | Drug Name |
|---|---|
| ***DB00500*** | Tolmetin |
| ***DB00501*** | Cimetidine |
| ***DB00502*** | Haloperidol |
| ***DB00518*** | Albendazole |
| ***DB00519*** | Trandolapril |
| ***DB00521*** | Carteolol |
| ***DB00530*** | Erlotinib |
| ***DB00532*** | Mephenytoin |
| ***DB00533*** | Rofecoxib |
| ***DB00535*** | Cefdinir |
| ***DB00537*** | Ciprofloxacin |
| ***DB00539*** | Toremifene |
| ***DB00540*** | Nortriptyline |
| ***DB00541*** | Vincristine |
| ***DB00542*** | Benazepril |
| ***DB00543*** | Amoxapine |
| ***DB00545*** | Pyridostigmine |
| ***DB00547*** | Desoximetasone |
| ***DB00548*** | Azelaic Acid |
| ***DB00549*** | Zafirlukast |
| ***DB00550*** | Propylthiouracil |
| ***DB00554*** | Piroxicam |
| ***DB00555*** | Lamotrigine |
| ***DB00558*** | Zanamivir |
| ***DB00559*** | Bosentan |
| ***DB00561*** | Doxapram |
| ***DB00563*** | Methotrexate |
| ***DB00564*** | Carbamazepine |
| ***DB00571*** | Propranolol |
| ***DB00572*** | Atropine |
| ***DB00573*** | Fenoprofen |
| ***DB00575*** | Clonidine |
| ***DB00580*** | Valdecoxib |
| ***DB00585*** | Nizatidine |
| ***DB00586*** | Diclofenac |
| ***DB00590*** | Doxazosin |
| ***DB00591*** | Fluocinolone Acetonide |
| ***DB00593*** | Ethosuximide |

| DBID | Drug Name |
|---|---|
| ***DB00594*** | Amiloride |
| ***DB00598*** | Labetalol |
| ***DB00602*** | Ivermectin |
| ***DB00603*** | Medroxyprogesterone |
| ***DB00605*** | Sulindac |
| ***DB00608*** | Chloroquine |
| ***DB00611*** | Butorphanol |
| ***DB00612*** | Bisoprolol |
| ***DB00615*** | Rifabutin |
| ***DB00619*** | Imatinib |
| ***DB00620*** | Triamcinolone |
| ***DB00621*** | Oxandrolone |
| ***DB00622*** | Nicardipine |
| ***DB00623*** | Fluphenazine |
| ***DB00624*** | Testosterone |
| ***DB00630*** | Alendronate |
| ***DB00631*** | Clofarabine |
| ***DB00633*** | Dexmedetomidine |
| ***DB00635*** | Prednisone |
| ***DB00640*** | Adenosine |
| ***DB00641*** | Simvastatin |
| ***DB00642*** | Pemetrexed |
| ***DB00647*** | Propoxyphene |
| ***DB00650*** | Leucovorin |
| ***DB00651*** | Dyphylline |
| ***DB00652*** | Pentazocine |
| ***DB00654*** | Latanoprost |
| ***DB00656*** | Trazodone |
| ***DB00659*** | Acamprosate |
| ***DB00661*** | Verapamil |
| ***DB00665*** | Nilutamide |
| ***DB00668*** | Epinephrine |
| ***DB00669*** | Sumatriptan |
| ***DB00672*** | Chlorpropamide |
| ***DB00673*** | Aprepitant |
| ***DB00674*** | Galantamine |
| ***DB00675*** | Tamoxifen |
| ***DB00678*** | Losartan |

| DBID | Drug Name |
|---|---|
| ***DB00679*** | Thioridazine |
| ***DB00680*** | Moricizine |
| ***DB00683*** | Midazolam |
| ***DB00685*** | Trovafloxacin |
| ***DB00687*** | Fludrocortisone |
| ***DB00690*** | Flurazepam |
| ***DB00691*** | Moexipril |
| ***DB00692*** | Phentolamine |
| ***DB00694*** | Daunorubicin |
| ***DB00695*** | Furosemide |
| ***DB00696*** | Ergotamine |
| ***DB00697*** | Tizanidine |
| ***DB00700*** | Eplerenone |
| ***DB00703*** | Methazolamide |
| ***DB00704*** | Naltrexone |
| ***DB00706*** | Tamsulosin |
| ***DB00708*** | Sufentanil |
| ***DB00710*** | Ibandronate |
| ***DB00712*** | Flurbiprofen |
| ***DB00714*** | Apomorphine |
| ***DB00715*** | Paroxetine |
| ***DB00720*** | Clodronate |
| ***DB00721*** | Procaine |
| ***DB00724*** | Imiquimod |
| ***DB00727*** | Nitroglycerin |
| ***DB00728*** | Rocuronium |
| ***DB00731*** | Nateglinide |
| ***DB00733*** | Pralidoxime |
| ***DB00734*** | Risperidone |
| ***DB00735*** | Naftifine |
| ***DB00740*** | Riluzole |
| ***DB00745*** | Modafinil |
| ***DB00747*** | Scopolamine |
| ***DB00749*** | Etodolac |
| ***DB00750*** | Prilocaine |
| ***DB00751*** | Epinastine |
| ***DB00753*** | Isoflurane |
| ***DB00754*** | Ethotoin |

| DBID | Drug Name |
|---|---|
| ***DB00757*** | Dolasetron |
| ***DB00758*** | Clopidogrel |
| ***DB00762*** | Irinotecan |
| ***DB00763*** | Methimazole |
| ***DB00764*** | Mometasone |
| ***DB00768*** | Olopatadine |
| ***DB00772*** | Malathion |
| ***DB00773*** | Etoposide |
| ***DB00774*** | Hydroflumethiazide |
| ***DB00775*** | Tirofiban |
| ***DB00776*** | Oxcarbazepine |
| ***DB00780*** | Phenelzine |
| ***DB00782*** | Propantheline |
| ***DB00783*** | Estradiol |
| ***DB00784*** | Mefenamic acid |
| ***DB00788*** | Naproxen |
| ***DB00790*** | Perindopril |
| ***DB00794*** | Primidone |
| ***DB00795*** | Sulfasalazine |
| ***DB00796*** | Candesartan |
| ***DB00798*** | Gentamicin |
| ***DB00799*** | Tazarotene |
| ***DB00800*** | Fenoldopam |
| ***DB00802*** | Alfentanil |
| ***DB00804*** | Dicyclomine |
| ***DB00806*** | Pentoxifylline |
| ***DB00807*** | Proparacaine |
| ***DB00808*** | Indapamide |
| ***DB00809*** | Tropicamide |
| ***DB00810*** | Biperiden |
| ***DB00811*** | Ribavirin |
| ***DB00813*** | Fentanyl |
| ***DB00814*** | Meloxicam |
| ***DB00818*** | Propofol |
| ***DB00819*** | Acetazolamide |
| ***DB00822*** | Disulfiram |
| ***DB00829*** | Diazepam |
| ***DB00831*** | Trifluoperazine |

| DBID | Drug Name |
|---|---|
| ***DB00834*** | Mifepristone |
| ***DB00835*** | Brompheniramine |
| ***DB00836*** | Loperamide |
| ***DB00838*** | Clocortolone |
| ***DB00839*** | Tolazamide |
| ***DB00841*** | Dobutamine |
| ***DB00842*** | Oxazepam |
| ***DB00843*** | Donepezil |
| ***DB00844*** | Nalbuphine |
| ***DB00850*** | Perphenazine |
| ***DB00851*** | Dacarbazine |
| ***DB00857*** | Terbinafine |
| ***DB00860*** | Prednisolone |
| ***DB00861*** | Diflunisal |
| ***DB00863*** | Ranitidine |
| ***DB00864*** | Tacrolimus |
| ***DB00868*** | Benzonatate |
| ***DB00869*** | Dorzolamide |
| ***DB00870*** | Suprofen |
| ***DB00871*** | Terbutaline |
| ***DB00872*** | Conivaptan |
| ***DB00873*** | Loteprednol |
| ***DB00876*** | Eprosartan |
| ***DB00881*** | Quinapril |
| ***DB00883*** | Isosorbide Dinitrate |
| ***DB00884*** | Risedronate |
| ***DB00887*** | Bumetanide |
| ***DB00889*** | Granisetron |
| ***DB00896*** | Rimexolone |
| ***DB00897*** | Triazolam |
| ***DB00898*** | Ethanol |
| ***DB00899*** | Remifentanil |
| ***DB00900*** | Didanosine |
| ***DB00903*** | Ethacrynic acid |
| ***DB00904*** | Ondansetron |
| ***DB00905*** | Bimatoprost |
| ***DB00906*** | Tiagabine |
| ***DB00907*** | Cocaine |

| DBID | Drug Name |
|---|---|
| ***DB00908*** | Quinidine |
| ***DB00909*** | Zonisamide |
| ***DB00910*** | Paricalcitol |
| ***DB00912*** | Repaglinide |
| ***DB00915*** | Amantadine |
| ***DB00918*** | Almotriptan |
| ***DB00920*** | Ketotifen |
| ***DB00921*** | Buprenorphine |
| ***DB00924*** | Cyclobenzaprine |
| ***DB00925*** | Phenoxybenzamine |
| ***DB00927*** | Famotidine |
| ***DB00929*** | Misoprostol |
| ***DB00933*** | Mesoridazine |
| ***DB00937*** | Diethylpropion |
| ***DB00938*** | Salmeterol |
| ***DB00949*** | Felbamate |
| ***DB00952*** | Naratriptan |
| ***DB00953*** | Rizatriptan |
| ***DB00959*** | Methylprednisolone |
| ***DB00960*** | Pindolol |
| ***DB00961*** | Mepivacaine |
| ***DB00962*** | Zaleplon |
| ***DB00963*** | Bromfenac |
| ***DB00964*** | Apraclonidine |
| ***DB00966*** | Telmisartan |
| ***DB00968*** | Methyldopa |
| ***DB00969*** | Alosetron |
| ***DB00973*** | Ezetimibe |
| ***DB00975*** | Dipyridamole |
| ***DB00978*** | Lomefloxacin |
| ***DB00979*** | Cyclopentolate |
| ***DB00980*** | Ramelteon |
| ***DB00981*** | Physostigmine |
| ***DB00983*** | Formoterol |
| ***DB00986*** | Glycopyrrolate |
| ***DB00988*** | Dopamine |
| ***DB00989*** | Rivastigmine |
| ***DB00990*** | Exemestane |

| DBID | Drug Name |
|---|---|
| ***DB00991*** | Oxaprozin |
| ***DB00992*** | Methyl aminolevulinate |
| ***DB00993*** | Azathioprine |
| ***DB00996*** | Gabapentin |
| ***DB00997*** | Doxorubicin |
| ***DB00998*** | Frovatriptan |
| ***DB00999*** | Hydrochlorothiazide |
| ***DB01001*** | Salbutamol |
| ***DB01005*** | Hydroxyurea |
| ***DB01006*** | Letrozole |
| ***DB01009*** | Ketoprofen |
| ***DB01012*** | Cinacalcet |
| ***DB01013*** | Clobetasol |
| ***DB01014*** | Balsalazide |
| ***DB01017*** | Minocycline |
| ***DB01018*** | Guanfacine |
| ***DB01019*** | Bethanechol |
| ***DB01023*** | Felodipine |
| ***DB01024*** | Mycophenolic acid |
| ***DB01029*** | Irbesartan |
| ***DB01030*** | Topotecan |
| ***DB01032*** | Probenecid |
| ***DB01035*** | Procainamide |
| ***DB01036*** | Tolterodine |
| ***DB01037*** | Selegiline |
| ***DB01039*** | Fenofibrate |
| ***DB01041*** | Thalidomide |
| ***DB01043*** | Memantine |
| ***DB01047*** | Fluocinonide |
| ***DB01050*** | Ibuprofen |
| ***DB01057*** | Echothiophate |
| ***DB01059*** | Norfloxacin |
| ***DB01062*** | Oxybutynin |
| ***DB01064*** | Isoproterenol |
| ***DB01067*** | Glipizide |
| ***DB01068*** | Clonazepam |
| ***DB01069*** | Promethazine |
| ***DB01073*** | Fludarabine |

| DBID | Drug Name |
|---|---|
| ***DB01076*** | Atorvastatin |
| ***DB01079*** | Tegaserod |
| ***DB01083*** | Orlistat |
| ***DB01085*** | Pilocarpine |
| ***DB01086*** | Benzocaine |
| ***DB01087*** | Primaquine |
| ***DB01088*** | Iloprost |
| ***DB01091*** | Butenafine |
| ***DB01095*** | Fluvastatin |
| ***DB01097*** | Leflunomide |
| ***DB01098*** | Rosuvastatin |
| ***DB01100*** | Pimozide |
| ***DB01101*** | Capecitabine |
| ***DB01104*** | Sertraline |
| ***DB01105*** | Sibutramine |
| ***DB01106*** | Levocabastine |
| ***DB01109*** | Heparin |
| ***DB01110*** | Miconazole |
| ***DB01114*** | Chlorpheniramine |
| ***DB01115*** | Nifedipine |
| ***DB01118*** | Amiodarone |
| ***DB01119*** | Diazoxide |
| ***DB01120*** | Gliclazide |
| ***DB01122*** | Ambenonium |
| ***DB01126*** | Dutasteride |
| ***DB01128*** | Bicalutamide |
| ***DB01129*** | Rabeprazole |
| ***DB01130*** | Prednicarbate |
| ***DB01132*** | Pioglitazone |
| ***DB01133*** | Tiludronate |
| ***DB01136*** | Carvedilol |
| ***DB01142*** | Doxepin |
| ***DB01143*** | Amifostine |
| ***DB01148*** | Flavoxate |
| ***DB01149*** | Nefazodone |
| ***DB01151*** | Desipramine |
| ***DB01156*** | Bupropion |
| ***DB01157*** | Trimetrexate |

| DBID | Drug Name |
|---|---|
| ***DB01158*** | Bretylium |
| ***DB01159*** | Halothane |
| ***DB01161*** | Chloroprocaine |
| ***DB01162*** | Terazosin |
| ***DB01165*** | Ofloxacin |
| ***DB01167*** | Itraconazole |
| ***DB01169*** | Arsenic trioxide |
| ***DB01173*** | Orphenadrine |
| ***DB01174*** | Phenobarbital |
| ***DB01177*** | Idarubicin |
| ***DB01182*** | Propafenone |
| ***DB01183*** | Naloxone |
| ***DB01184*** | Domperidone |
| ***DB01185*** | Fluoxymesterone |
| ***DB01186*** | Pergolide |
| ***DB01189*** | Desflurane |
| ***DB01193*** | Acebutolol |
| ***DB01194*** | Brinzolamide |
| ***DB01195*** | Flecainide |
| ***DB01196*** | Estramustine |
| ***DB01197*** | Captopril |
| ***DB01198*** | Zopiclone |
| ***DB01200*** | Bromocriptine |
| ***DB01202*** | Levetiracetam |
| ***DB01203*** | Nadolol |
| ***DB01204*** | Mitoxantrone |
| ***DB01205*** | Flumazenil |
| ***DB01206*** | Lomustine |
| ***DB01210*** | Levobunolol |
| ***DB01214*** | Metipranolol |
| ***DB01215*** | Estazolam |
| ***DB01216*** | Finasteride |
| ***DB01217*** | Anastrozole |
| ***DB01218*** | Halofantrine |
| ***DB01219*** | Dantrolene |
| ***DB01220*** | Rifaximin |
| ***DB01221*** | Ketamine |
| ***DB01222*** | Budesonide |

| DBID | Drug Name |
|---|---|
| ***DB01223*** | Aminophylline |
| ***DB01224*** | Quetiapine |
| ***DB01226*** | Mivacurium |
| ***DB01229*** | Paclitaxel |
| ***DB01233*** | Metoclopramide |
| ***DB01234*** | Dexamethasone |
| ***DB01236*** | Sevoflurane |
| ***DB01238*** | Aripiprazole |
| ***DB01241*** | Gemfibrozil |
| ***DB01242*** | Clomipramine |
| ***DB01247*** | Isocarboxazid |
| ***DB01248*** | Docetaxel |
| ***DB01250*** | Olsalazine |
| ***DB01254*** | Dasatinib |
| ***DB01258*** | Aliskiren |
| ***DB01260*** | Desonide |
| ***DB01261*** | Sitagliptin |
| ***DB01267*** | Paliperidone |
| ***DB01268*** | Sunitinib |
| ***DB01273*** | Varenicline |
| ***DB01275*** | Hydralazine |
| ***DB01276*** | Exenatide |
| ***DB01278*** | Pramlintide |
| ***DB01280*** | Nelarabine |
| ***DB01291*** | Pirbuterol |
| ***DB01306*** | Insulin Aspart |
| ***DB01320*** | Fosphenytoin |
| ***DB01327*** | Cefazolin |
| ***DB01337*** | Pancuronium |
| ***DB01340*** | Cilazapril |
| ***DB01356*** | Lithium |
| ***DB01364*** | Ephedrine |
| ***DB01367*** | Rasagiline |
| ***DB01373*** | Calcium |
| ***DB01378*** | Magnesium |
| ***DB01393*** | Bezafibrate |
| ***DB01394*** | Colchicine |
| ***DB01399*** | Salsalate |

| DBID | Drug Name | DBID | Drug Name | DBID | Drug Name |
|---|---|---|---|---|---|
| ***DB01400*** | Neostigmine | ***DB01621*** | Pipotiazine | ***DB06209*** | Prasugrel |
| ***DB01406*** | Danazol | ***DB01623*** | Thiothixene | ***DB06228*** | Rivaroxaban |
| ***DB01409*** | Tiotropium | ***DB02300*** | Calcipotriol | ***DB06274*** | Alvimopan |
| ***DB01410*** | Ciclesonide | ***DB04835*** | Maraviroc | ***DB06287*** | Temsirolimus |
| ***DB01427*** | Amrinone | ***DB04839*** | Cyproterone | ***DB06335*** | Saxagliptin |
| ***DB01558*** | Bromazepam | ***DB04844*** | Tetrabenazine | ***DB06695*** | Dabigatran etexilate |
| ***DB01577*** | Methamphetamine | ***DB04845*** | Ixabepilone | ***DB06698*** | Betahistine |
| ***DB01586*** | Ursodeoxycholic acid | ***DB04861*** | Nebivolol | ***DB06699*** | Degarelix |
| ***DB01591*** | Solifenacin | ***DB04868*** | Nilotinib | ***DB06700*** | Desvenlafaxine |
| ***DB01595*** | Nitrazepam | ***DB04896*** | Milnacipran | ***DB06702*** | Fesoterodine |
| ***DB01611*** | Hydroxychloroquine | ***DB04930*** | Permethrin | ***DB06710*** | Methyltestosterone |
| ***DB01612*** | Amyl Nitrite | ***DB05246*** | Methsuximide | ***DB06711*** | Naphazoline |
| ***DB01618*** | Molindone | ***DB05271*** | Rotigotine | ***DB06802*** | Nepafenac |

Drugs were obtained from the DrugBank database[2], and their side effects were collected from the SIDER database[3].

**Table 2 │ The keywords used in the filtering of the DrugBank database and their occurrences**

| Keyword | Mark | Occurrences |
|---|---|---|
| „cancer”/ „lymphoma”/ „carcinoma”/ „leukemia”/ „tumor” | Anti-cancer | 172 |
| „colon”/ „colorectal”/ „carcinoma”/ „cancer”/ „tumor” | Anti-colorectal cancer | 11 |
| „diabetes mellitus” | Anti-diabetes | 36 |

The keywords are listed which were used in the filtering of the DrugBank database[2] and their occurrences is noted. The plus sign (+) represents the “AND” logical operator, the slash (/) represents the “OR” logical operator.

**Table 3 | Drugs obtained from the DrugBank database, which are used in the treatment of colorectal cancer and have no reported side effects in the SIDER database and their target proteins**

| DrugBank ID | Drug Name | Drug Target Proteins |
|---|---|---|
| ***DB00002*** | Cetuximab | O75015, P00533, P00736, P02745, P02746, P02747, P09871, P12314, P12318, P31994 |
| ***DB00112*** | Bevacizumab | O75015, P00736, P02745, P02746, P02747, P12314, P12318, P31994 |
| ***DB00113*** | Arcitumomab | P13688 |
| ***DB00544*** | Fluorouracil | P04818 |
| ***DB00848*** | Levamisole | P10696, P32297 |
| ***DB01269*** | Panitumumab | P00533 |
| ***DB01873*** | Epothilone D | P04350, P07437, P68363, P68366, P68371, Q13509, Q13748, Q71U36, Q9BQE3, Q9H4B7, Q9NY65 |

Drugs and their targets were obtained from the DrugBank database[2]. Only those drugs were selected, which are used in the treatment of colorectal cancer and have no reported side effects in the SIDER database[3]. Target proteins for each drug were identified by their UniProt ID[9].

**Table 4 | Drugs obtained from the DrugBank database, which are used in the treatment of colorectal cancer and have known side effects in the SIDER database and their target proteins**

| **Drugbank ID** | **Drug Name** | **Drug Target Proteins** |
|---|---|---|
| ***DB00650*** | Leucovorin | P04818 |
| ***DB00762*** | Irinotecan | P11387 |
| ***DB01101*** | Capecitabine | P04818 |
| ***DB01157*** | Trimetrexate | P00374 |

Drugs and their targets were obtained from the DrugBank database[2]. Only those drugs were selected, which are used in the treatment of colorectal cancer and have known side effects in the SIDER database[3]. Target proteins for each drug were identified by their UniProt ID[9].

**Table 5 | Drugs obtained from the DrugBank database, which are used in the treatment of type 2 diabetes and have no reported side effects in the SIDER database and their target proteins**

| DrugBank ID | Drug Name | Drug Target Proteins |
|---|---|---|
| ***DB00030*** | Insulin recombinant | P06213, P06400, P07339, P08069, P14735, P16519, P16870, P29120, P48745, P98164, Q16270, Q96C24 |
| ***DB00071*** | Insulin, porcine | P01906, P06213, P06400, P07339, P08069, P14735, P16519, P16870, P29120, P48745, P98164, Q16270, Q96C24 |
| ***DB00414*** | Acetohexamide | P48048 |
| ***DB00722*** | Lisinopril | P12821, Q9BYF1 |
| ***DB00914*** | Phenformin | Q13131, Q15842 |
| ***DB01124*** | Tolbutamide | P48048, Q09428 |
| ***DB01251*** | Gliquidone | Q09428, Q15842 |
| ***DB01289*** | Glisoxepide | Q09428, Q15842 |
| ***DB01307*** | Insulin Detemir | P06213 |
| ***DB01309*** | Insulin Glulisine | P06213 |
| ***DB01382*** | Glycodiazine | P48048, Q09428 |
| ***DB04876*** | Vildagliptin | P27487 |
| ***DB06655*** | Liraglutide | P43220 |

Drugs and their targets were obtained from the DrugBank database[2]. Only those drugs were selected, which are used in the treatment of type 2 diabetes and have no reported side effects in the SIDER database[3]. Target proteins for each drug were identified by their UniProt ID[9].

**Table 6 │ Drugs obtained from the DrugBank database, which are used in the treatment of type 2 diabetes and have known side effects in the SIDER database and their target proteins**

| Drugbank ID | Drug Name | Drug Target Proteins |
|---|---|---|
| ***DB00046*** | Insulin Lispro | P06213, P08069 |
| ***DB00047*** | Insulin Glargine | P06213, P08069 |
| ***DB00178*** | Ramipril | P12821 |
| ***DB00197*** | Troglitazone | O60488, P05121, P11474, P37231, P62508, Q99808 |
| ***DB00222*** | Glimepiride | P48048, Q09428, Q14654 |
| ***DB00412*** | Rosiglitazone | O60488, P37231 |
| ***DB00491*** | Miglitol | P10253, Q14697, Q8TET4 |
| ***DB00492*** | Fosinopril | P12821 |
| ***DB00519*** | Trandolapril | P12821 |
| ***DB00731*** | Nateglinide | P37231, Q09428 |
| ***DB00834*** | Mifepristone | P04150, P06401 |
| ***DB00839*** | Tolazamide | P48048 |
| ***DB00881*** | Quinapril | P12821 |
| ***DB00912*** | Repaglinide | P37231, Q09428 |
| ***DB00966*** | Telmisartan | P30556, P37231 |
| ***DB01067*** | Glipizide | P37231, Q09428 |
| ***DB01132*** | Pioglitazone | P37231 |
| ***DB01261*** | Sitagliptin | P27487 |
| ***DB01276*** | Exenatide | P43220 |
| ***DB01278*** | Pramlintide | O60894, O60895, O60896 |
| ***DB01306*** | Insulin Aspart | P06213 |
| ***DB01393*** | Bezafibrate | P37231, Q03181, Q07869 |
| ***DB06335*** | Saxagliptin | P27487 |

Drugs and their targets were obtained from the DrugBank database[2]. Only those drugs were selected, which are used in the treatment of type 2 diabetes and have known side effects in the SIDER database[3]. Target proteins for each drug were identified by their UniProt ID[9].

**Table 7 | Mutated genes in colorectal cancer and their corresponding proteins**

| Gene name | Protein identifier |
|---|---|
| ***AKT1*** | P31749 |
| ***APC*** | P25054 |
| ***BRAF*** | P15056 |
| ***CTNNB1*** | P35222 |
| ***EP300*** | Q09472 |
| ***FBXW7*** | Q969H0 |
| ***KRAS*** | P01116 |
| ***MADH4*** | Q13485 |
| ***MAP2K4*** | P45985 |
| ***MDM2*** | Q00987 |
| ***MLH1*** | P40692 |
| ***MSH2*** | P43246 |
| ***MSH6*** | P52701 |
| ***PIK3CA*** | P42336 |
| ***PIK3R1*** | P27986 |
| ***TCF7L2*** | Q9NQB0 |
| ***TP53*** | P04637 |
| ***VTI1A*** | Q96AJ9 |

The 18 mutated genes in colorectal cancer were obtained from the Cancer Gene Census[10] and the proteins coded by them were mapped by PICR[11].

**Table 8 | Mutated genes in type 2 diabetes and their corresponding proteins**

| Gene name | Protein identifier |
|---|---|
| ***ABCC8*** | Q54P13 |
| ***CAPN10*** | Q9HC96 |
| ***HNF1B*** | Q91910 |
| ***GCGR*** | P30082 |
| ***TCF7L2*** | Q9NQB0* |
| ***PPARG*** | O18924 |
| ***KCNJ11*** | O02822 |
| ***WFS1*** | P56695 |
| ***HNF1B*** | Q91910 |
| ***SLC30A8*** | Q5I020 |
| ***HHEX*** | D2KQB0 |
| ***CDKAL1*** | Q5VV42* |
| ***IGF2BP2*** | Q9Y6M1* |
| ***CDKN2A*** | O77617 |
| ***CDKN2B*** | P42772* |
| ***FTO*** | Q9C0B1* |
| ***JAZF1*** | Q80ZQ5 |
| ***CDC123*** | A6R687 |
| ***CAMK1D*** | Q8IU85* |
| ***TSPAN8*** | Q2KIS9 |
| ***LGR5*** | Q9Z1P4 |
| ***THADA*** | A8C752 |
| ***ADAMTS9*** | Q9P2N4 |
| ***NOTCH2*** | Q04721* |
| ***KCNQ1*** | P51787* |
| ***IRS1*** | Q28224 |
| ***MTNR1B*** | Q8CIQ6 |
| ***PROX1*** | P48437 |
| ***GCKR*** | Q07071 |
| ***ADCY5*** | P30803 |
| ***UBE2E2*** | Q96LR5* |
| ***BCL11A*** | Q9H165* |
| ***GCKR*** | Q07071 |
| ***DGKB*** | Q9Y6T7* |
| ***TMEM195*** | A0JPQ8 |
| ***C2CD4B*** | A6NLJ0 |
| ***KLF14*** | Q9ESX2 |
| ***ZBED3*** | Q96IU2 |
| ***TP53INP1*** | Q96A56* |
| ***CHCHD9*** | Q5T1J5 |
| ***CENTD2*** | Q4LDD4 |
| ***HMGA2*** | P52926* |
| ***HNF1A*** | Q90867 |
| ***PRC1*** | Q94JQ6 |
| ***ZFAND6*** | Q9DCH6 |
| ***DUSP9*** | Q99956* |

The 46 mutated genes in type 2 diabetes were obtained from the article of Parchwani et al.[12] and the proteins coded by them were mapped by PICR[10]. From the 46 proteins listed here only 14 were contained in the human interactome constructed from the STRING database[1]; those are marked with an asterisk (*) in the Table.

**Table 9 | Average human interactome centralities of target proteins of drugs against colorectal cancer and type 2 diabetes**

| Centrality type | Drug targets without side effects | | | Drug targets with side effects | | |
|---|---|---|---|---|---|---|
| | Colorectal cancer | Type 2 diabetes | Statistical difference | Colorectal cancer | Type 2 diabetes | Statistical difference |
| **Degree** (number of neighbours) | 24.50 | 13.00 | 0.203 | 40.00 | 34.00 | 0.941 |
| **Closeness centrality** (1/edge) | 0.305 | 0.295 | 0.330 | 0.301 | 0.292 | 0.572 |
| **Betweenness centrality** (fraction of shortest paths passing through the node) | 1.46E-4 | 5.76E-4 | 0.601 | 3.39E-4 | 1.28E-4 | 0.944 |

The table shows the medians of the centralities of target proteins of drugs against colorectal cancer and type 2 diabetes without or with reported side effects (the results were very similar, if instead of medians we used the arithmetic means; data not shown). Centrality values were calculated with the Pajek programme[6]. The human interactome containing 12,439 proteins and 174,666 edges was built from the STRING database[1], 1,726 human drug targets were obtained from the DrugBank database[2], and the proteins were labelled by their UniProt ID[9]. 99,423 drug-side effect pairs were taken from the SIDER database[3]. Statistical analysis was performed using the Wilcoxon rank sum (Mann-Whitney) test function of the R package[5].

**Table 10 | Average network distance between drug targets without known side effects used in the treatment of colorectal cancer and colorectal cancer-associated proteins**

| **UniProt ID of colorectal cancer drug targets without side effects** | **Average network distance from colorectal cancer-related proteins** (edges) |
|---|---|
| *O75015* | 2.500 |
| *P00533* | 1.722 |
| *P00736* | 2.722 |
| *P02745* | 2.889 |
| *P02746* | 3.000 |
| *P02747* | 3.000 |
| *P04350* | 2.278 |
| *P07437* | 2.167 |
| *P09871* | 3.000 |
| *P10696* | 3.222 |
| *P12314* | 2.722 |
| *P12318* | 2.444 |
| *P13688* | 2.444 |
| *P31994* | 2.500 |
| *P32297* | 3.056 |
| *P68363* | 2.111 |
| *P68366* | 2.000 |
| *P68371* | 2.444 |
| *Q13509* | 2.722 |
| *Q13748* | 2.111 |
| *Q71U36* | 2.111 |
| *Q9BQE3* | 2.389 |
| *Q9H4B7* | 2.778 |
| *Q9NY65* | 2.333 |
| ***Mean network distance of drug targets*** | **2.528** |
| ***Mean network distance of randomly selected proteins*** | **3.316** |

The table shows the average network distance between drug targets without known side effects used in the treatment of colorectal cancer and colorectal cancer-related proteins. The total number of drug targets without known side effects used in the treatment of colorectal cancer was 24; the total number of colorectal cancer-related proteins was 18. Average network distances were calculated as shortest paths using the Pajek programme[6]. Proteins were labelled by their UniProt ID[9]. The human interactome containing 12,439 proteins and 174,666 edges was built from the STRING database[1], 1,726 human drug targets were obtained from the DrugBank database[2] and 99,423 drug-side effect pairs were taken from the SIDER database[3]. Colorectal cancer-related proteins were obtained from the Cancer Gene Census database[10]. Average network distances between colorectal cancer-related proteins and at least 50 randomly selected samples of 24 proteins each were calculated, and the statistical difference in their mean values compared to the average network distance of the 24 drug targets listed above was tested using the one-way ANOVA (Analysis of Variance) with linear model fit function of the R package[5]. There was no statistically significant difference between the mean values of the drug targets without known side effects and the random samples, $F=0.8807$, $p=0.7078$.

**Table 11 | Average network distance between drug targets with known side effects used in the treatment of colorectal cancer and colorectal cancer-associated proteins**

| UniProt ID of colorectal cancer drug targets with side effects | Average network distance from colorectal cancer-related proteins (edges) |
|---|---|
| *P00374* | 2.500 |
| *P04818* | 2.556 |
| *P11387* | 2.111 |
| ***Mean network distance of drug targets*** | **2.389** |
| ***Mean network distance of randomly selected proteins*** | **3.240** |

The table shows the average network distance between drug targets with known side effects used in the treatment of colorectal cancer and colorectal cancer-related proteins. The total number of drug targets with known side effects used in the treatment of colorectal cancer was 3; the total number of colorectal cancer-related proteins was 18. Average network distances were calculated as shortest paths using the Pajek programme[6]. Proteins were labelled by their UniProt ID[9]. The human interactome containing 12,439 proteins and 174,666 edges was built from the STRING database[1], 1,726 human drug targets were obtained from the DrugBank database[2] and 99,423 drug-side effect pairs were taken from the SIDER database[3]. Colorectal cancer-related proteins were obtained from the Cancer Gene Census database[10]. Average network distances between colorectal cancer related proteins and at least 50 randomly selected samples of 3 proteins each were calculated, and the statistical difference in their mean values compared to the average network distance of the 3 drug targets listed above was tested using the one-way ANOVA (Analysis of Variance) with linear model fit function of the R package[5]. There was no statistically significant difference between the mean values of the drug targets with known side effects and the random samples, $F=1.223$, $p=0.1951$.

**Table 12 | Average network distance between drug targets without known side effects used in the treatment of type 2 diabetes and diabetes-associated proteins**

| UniProt ID of type 2 diabetes drug targets without side effects | Average network distance from diabetes-related proteins (edges) |
|---|---|
| *P01906* | *3.786* |
| *P06400* | *2.286* |
| *P07339* | *3.000* |
| *P14735* | *3.214* |
| *P16519* | *3.786* |
| *P16870* | *3.286* |
| *P29120* | *3.143* |
| *P48745* | *3.214* |
| *P98164* | *3.000* |
| *Q13131* | *2.929* |
| *Q15842* | *3.714* |
| *Q16270* | *3.143* |
| *Q96C24* | *3.500* |
| *Q9BYF1* | *3.500* |
| ***Mean network distance of drug targets*** | ***3.250*** |
| ***Mean network distance of randomly selected proteins*** | ***3.413*** |

The table shows the average network distance between drug targets without known side effects used in the treatment of type 2 diabetes and diabetes-related proteins. The total number of drug targets without known side effects used in the treatment of type 2 diabetes was 14; the total number of type 2 diabetes-related proteins contained in the human interactome was 14. Average network distances were calculated as shortest paths using the Pajek programme[6]. Proteins were labelled by their UniProt ID[9]. The human interactome containing 12,439 proteins and 174,666 edges was built from the STRING database[1], 1,726 human drug targets were obtained from the DrugBank database[2] and 99,423 drug-side effect pairs were taken from the SIDER database[3]. Type 2 diabetes-related proteins were obtained from the article of Parchwani et al.[12]. Average network distances between type-2 diabetes related proteins and at least 50 randomly selected samples of 14 proteins each were calculated, and the statistical difference in their mean values compared to the average network distance of the 14 drug targets listed above was tested using the one-way ANOVA (Analysis of Variance) with linear model fit function of the R package[5]. There was no statistically significant difference between the mean values of the drug targets without known side effects and the random samples, $F=0.7867$, $p=0.8547$.

**Table 13 │Average network distance between drug targets with known side effects used in the treatment of type 2 diabetes and diabetes-associated proteins**

| **UniProt ID of type 2 diabetes drug targets with side effects** | **Average network distance from diabetes-related proteins** (edges) |
|---|---|
| *O60488* | 3.643 |
| *O60894* | 3.857 |
| *O60895* | 3.857 |
| *O60896* | 3.429 |
| *P04150* | 2.429 |
| *P05121* | 2.857 |
| *P06213* | 2.643 |
| *P06401* | 2.500 |
| *P08069* | 2.571 |
| *P10253* | 3.786 |
| *P11474* | 3.000 |
| *P12821* | 3.786 |
| *P27487* | 3.500 |
| *P30556* | 3.000 |
| *P37231* | 2.643 |
| *P43220* | 3.071 |
| *P48048* | 3.214 |
| *P62508* | 3.071 |
| *Q03181* | 2.857 |
| *Q07869* | 2.714 |
| *Q09428* | 3.500 |
| *Q14654* | 3.286 |
| *Q14697* | 3.357 |
| *Q8TET4* | 3.929 |
| *Q99808* | 4.357 |
| ***Mean network distance of drug targets*** | **3.234** |
| ***Mean network distance of randomly selected proteins*** | **3.443** |

The table shows the average network distance between drug targets with known side effects used in the treatment of type 2 diabetes and diabetes-related proteins. The total number of drug targets with known side effects used in the treatment of type 2 diabetes was 25; the total number of type 2 diabetes-related proteins contained in the human interactome was 14. Average network distances were calculated as shortest paths using the Pajek programme[6]. Proteins were labelled by their UniProt ID[9]. The human interactome containing 12,439 proteins and 174,666 edges was built from the STRING database[1], 1,726 human drug targets were obtained from the DrugBank database[2] and 99,423 drug-side effect pairs were taken from the SIDER database[3]. Type 2 diabetes-related proteins were obtained from the article of Parchwani et al.[12]. Average network distances between type-2 diabetes related proteins and at least 50 randomly selected samples of 25 proteins each were calculated, and the statistical difference in their mean values compared to the average network distance of the 25 drug targets listed above was tested using the one-way ANOVA (Analysis of Variance) with linear model fit function of the R package[5]. There was no statistically significant difference between the mean values of the drug targets with known side effects and the random samples, $F= 0.9021$, $p= 0.6677$.

## Supplementary References

1. Franceschini, A. *et al.* STRING v9.1: protein-protein interaction networks, with increased coverage and integration. *Nucleic Acids Res.* **41,** D808–D815 (2012).
2. Knox, C. *et al.* DrugBank 3.0: a comprehensive resource for 'omics' research on drugs. *Nucleic Acids Res.* **39,** D1035–1041 (2011).
3. Kuhn, M., Campillos, M., Letunic, I., Jensen, L. J. & Bork, P. A side effect resource to capture phenotypic effects of drugs. *Mol. Syst. Biol.* **6,** 343 (2010).
4. Szalay, K. Z. & Csermely, P. Perturbation centrality and Turbine: A novel centrality measure obtained using a versatile network dynamics tool. *PLoS ONE* **8,** e78059 (2013).
5. R Core Team. *R: A language and environment for statistical computing.* (R Foundation for Statistical Computing, 2013). at <http://www.R-project.org/>
6. Vladimir Bagatelj & Andrej Mrvar. in *Graph drawing software* (eds. Jünger, M. & Mutzel, P.) 77–103 (Springer, 2003). at <http://pajek.imfm.si/>
7. The Inkscape Team. *Inkscape*. (2014). at <http://inkscape.org>
8. Shannon, P. *et al.* Cytoscape: a software environment for integrated models of biomolecular interaction networks. *Genome Res.* **13,** 2498–2504 (2003).
9. The UniProt Consortium. Reorganizing the protein space at the Universal Protein Resource (UniProt). *Nucleic Acids Res.* **40,** D71–D75 (2012).
10. Forbes, S. A. *et al.* COSMIC: mining complete cancer genomes in the Catalogue of Somatic Mutations in Cancer. *Nucleic Acids Res.* **39,** D945–D950 (2010).
11. Wein, S. P. *et al.* Improvements in the Protein Identifier Cross-Reference service. *Nucleic Acids Res.* **40,** W276–280 (2012).
12. Parchwani, D., Murthy, S., Upadhyah, A. & Patel, D. Genetic factors in the etiology of type 2 diabetes: linkage analyses, candidate gene association, and genome-wide association – still a long way to go! *Natl. J. Physiol. Pharm. Pharmacol.* **3,** 57 (2013).